\documentclass{article} 

\usepackage{latexsym,amssymb,amsfonts,amsmath,amsthm}
\usepackage[dvips]{graphicx}

\title{QUANTUM GRAPH WALKS II \\
Quantum walks on graph coverings}
\author{Yusuke HIGUCHI \\ 
Mathematics Laboratories, College of Arts and Sciences, \\ 
Showa University, 4562 Kamiyoshida, Fujiyoshida,
Yamanashi 403-0005, Japan \\ 
Norio KONNO \\
Department of Applied Mathematics, \\
Faculty of Engineering, Yokohama National University \\
Hodogaya, Yokohama 240-8501, Japan \\
Iwao SATO \\ 
Oyama National College of Technology, 
Oyama, Tochigi 323-0806, Japan \\
Etsuo SEGAWA \\
Graduate School of Information Sciences, Tohoku University \\
Sendai 980-8579, Japan.} 

\begin{document}
 \maketitle

\clearpage

\begin{abstract}
We give a new determinant expression for the characteristic polynomial of 
the bond scattering matrix of a quantum graph $G$. 
Also, we give a decomposition formula for the characteristic polynomial of 
the bond scattering matrix of a regular covering of $G$. 
Furthermore, we define an $L$-function of $G$, and give a determinant 
expression of it.
As a corollary, we express the characteristic polynomial of the bond scattering 
matrix of a regular covering of $G$ by means of its $L$-functions. 
As an application, we introduce three types of quantum graph walks, 
and treat their relation. 
\end{abstract}

\vspace{5mm}

{\bf 2000 Mathematical Subject Classification}: 81Q10, 05C50, 05C60, 81P68. \\
{\bf Key words and phrases} : quantum graph, scattering matrix, quantum walk, 
graph covering, $L$-function 

\vspace{5mm}

The contact author for correspondence:

\clearpage

\section{Introduction}

A quantum graph identifies edges of an ordinary graph with 
closed intervals generating a metric graph, and has an operator acting on 
functions defined on the collection of intervals. 
The review and book on quantum graphs are 
Exner and \v{S}eba [8], Kuchment [25], Gnutzmann and Smilansky [11], for examples. 

One of interest on quantum graphs is the spectral question of quantum graphs. 
This is approached through a trace formula. 
The first graph trace formula was derived by Roth [29]. 
Kottos and Smilansy [24] introduced a contour integral approach to the trace formula 
starting with a secular equation based on the scattering matrix of plane-waves on the graph. 
Solutions of the secular equation corresponds to the points in the spectrum of the quantum graph. 

Trace formulas express spectral functions like the density of states or heat kernel 
as sums over periodic orbits on the graph. 
This fact is related to the Ihara zeta function. 
furthermore, the spectral determinant of the Laplacian on a quantum graph 
is closely related to the Ihara zeta function of a graph (see [5,6,13,14]). 
Smilansky [32] considered spectral zeta functions and trace formulas 
for (discrete) Laplacians on ordinary graphs, and expressed 
some determinant on the bond scattering matrix of a graph $G$ 
by using the characteristic polynomial of its Laplacian.

As a quantum counterpart of the classical random walk, 
the quantum walk has recently attracted much attention for various fields. 
The review and book on quantum walks are Ambainis [1], Kempe [19], Kendon [20], 
Konno [21], Venegas-Andraca [39], for examples. 

In 1988, Gudder defined discrete-time quantum walk on a graph 
from the view point of quantum measure introduced as a quantum 
analogue of probability measure in his book [12]. 
The Grover walk on a graph was formulated in [41].  
We can see that there are many applications of the Grover walk 
to quantum spatial search algorithms in the review by Ambainis [1], 
for example. 
As a generalization of the Grover walk, Szegedy [37] introduced the Szegedy walk 
on a graph related to a transition matrix of a random walk on the same graph. 

Recently, the relation between quantum graphs and quantum walks on graphs 
are pointed out (see [31,38]).  
Higuchi, Konno, Sato and Segawa [16] introduced a quantum walks related quantum graphs. 
As a sequential work of this manuscript and [16], we show the 
relationship between a quantum walk and a scattering amplitude via 
discrete Laplacian in [17]. 

Zeta functions of graphs were originally defined for regular graphs 
by Ihara [18]. 
This is the Ihara zeta function of a graph. 
In [18], he showed that their reciprocals are explicit polynomials. 
A zeta function of a regular graph $G$ associated with a unitary 
representation of the fundamental group of $G$ was developed by 
Sunada [35,36]. 
Hashimoto [15] treated multivariable zeta functions of bipartite graphs. 
Bass [4] generalized Ihara's result on the zeta function of 
a regular graph to an irregular graph and showed that its reciprocal is 
again a polynomial. 
A decomposition formula for the Ihara zeta function of a regular covering of a graph 
was obtained by Stark and Terras [34], and independently, Mizuno and Sato [27]. 

The discrete-time quantum walk on a graph is closely related to the Ihara zeta function 
of a graph. 
Ren et al. [28] found an interesting relation between the Ihara zeta function and 
the discrete-time quantum walk on a graph, and showed that the positive support of 
the transition matrix of the discrete-time quantum walk is equal to the Perron-Frobenius operator 
(the edge matrix) related to the Ihara zeta function. 
Konno and Sato [22] gave the characteristic polynomials of the support of the transition
matrix of the discrete-time quantum walk and its positive support, and so obtained 
the other proofs of the results on spectra for them by Emms et al. [7]. 

In this paper, we present a new determinant expression for the scattering matrix of 
a quantum graph. 
In Section 2, we state a short review on quantum graphs. 
We consider the Schr\"{o}dinger equation and the boundary conditions of a quantum graph 
from a view point of arcs (oriented edges) of the graph under Ref. [16], and present two types of 
the scattering matrix of a quantum graph. 
In Section 3, we treat a quantum walk on a graph, and 
discuss the relation between %three 
four quantum graph walks induced by a quantum graph. 
We clarify that these walks are in spatial and temporal reversal relation. 
In Section 4, we present a new determinant expression for the characteristic polynomial of 
the scattering matrix of a quantum graph by using the method of Watanabe and Fukumizu [40].
In Section 5, we give a formulation for the Schr\"{o}dinger equation and the boundary conditions 
of a regular covering of a quantum graph, and propose a type of the scattering matrix of 
a quantum graph whose base graph is a regular covering of a graph. 
Furthermore, we give a decomposition formula for the characteristic polynomial of 
the scattering matrix of a regular covering. 
In Section 6, we define an $L$-function of a graph and give a determinant 
expression for it. 
As a corollary, we express the determinant for the characteristic polynomial of 
the scattering matrix of a regular covering as a product of $L$-functions. 
In Section 7, we express the above $L$-function of a graph by using the Euler 
product.

\section{Scattering matrix of a quantum graph} 

We present a review on a quantum graph. 

Graphs treated here are finite.
Let $G$ be a connected graph (possibly with multiple edges and loops) 
with the set $V(G)$ of vertices and the set $E(G)$ of unoriented edges. 
We write  $uv$ for an edge joining two vertices $u$ and $v$. 
For $uv \in E(G)$, an arc $(u,v)$ is the oriented edge from $u$ to $v$. 
Set $D(G)= \{ (u,v),(v,u) \mid uv \in E(G) \} $. 
For $e=(u,v) \in D(G)$, set $u=o(e)$ and $v=t(e)$. 
Furthermore, let $e^{-1}=(v,u)$ be the {\em inverse} arc of $e=(u,v)$.

Let $G$ be a connected graph with $V(G)= \{ 1 , \ldots , n \} $ 
and $D(G)= \{ e_1 , \ldots , e_m , e^{-1}_1 , \ldots , e^{-1}_m \} $. 
Arrange vertices of $G$ as follows: $1<2< \cdots <n$. 
Furthermore, let $d_j = \deg j, j \in V(G)$. 
For each edge $ij \in E(G)$, let $L_{ij} $ and $A_{ij} $ be the length and 
the vector potential of $ij$, respectively. 
If $ij \in E(G)$, then assign a variable $x$ in the interval $[0, L_{ij} ]$ 
such that $x=0$ and $x=L_{ij} $ corresponds to $i$ and $j$, respectively, 
and an intermediate point $z$ of $ij$ corresponds to the distance between $i$ and $z$. 

For $e=(i,j) \in D(G)$, set 
\[
L_e = L_{ij} , 
A_{e} =\left\{
\begin{array}{ll}
A_{ij} & \mbox{if $i<j$, } \\
- A_{ij} & \mbox{if $i>j$. }
\end{array}
\right. 
\] 
Note that 
\[
L_e = L_{e^{-1}} , \  \  A_{e^{-1}} =- A_e . 
\]

Let $e=(j,l) \in D(G)$. 
Then the Schr\"{o}dinger equation for $e$ is given by 
\begin{equation}
\left(-{\bf i} {\rm \frac{d}{dx} } + A_e \right)^2 \Psi {}_e (x)= k^2 \Psi {}_e (x) 
\end{equation}
under the following three conditions: 
\begin{enumerate} 
\item $ \Psi {}_e (x)= \Psi {}_{e^{-1} } (L_{jl} -x)$;  
\item {\em The continuity}: $\Psi {}_e (0)= \phi {}_j $ and $\Psi {}_e (L_{jl})= \phi {}_l $; 
\item {\em The cuurent conservation}: 
\[
\sum_{o(e)=j} \left(-{\bf i} {\rm \frac{d}{dx} } + A_e \right) \Psi {}_e (x) \bigg|_{x=0} =-{\bf i} \lambda {}_j \phi {}_j , 
\forall j \in V(G) ,  
\] 
where $( \phi {}_1 , \ldots , \phi {}_n ) \in {\bf C}^n $. 
\end{enumerate} 

The solution of (1) is given by 
\begin{equation}
\Psi {}_e (x)=( a_e e^{-{\bf i}kx} + b_e e^{{\bf i}kx}) e^{-{\bf i}A_e x} , {\bf i}= \sqrt{-1} . 
\end{equation}
By condition 1, we have 
\begin{equation}
a_e = b_{e^{-1}} e^{{\bf i}L_e (k+ A_e )} \  and \  b_e = a_{e^{-1}} e^{-{\bf i}L_e (k- A_e )} . 
\end{equation}
By condition 2, we have 
\begin{equation}
a_{e_1} +b_{e_1 } =a_{e_2} +b_{e_2 } = \cdots = a_{e_{d_j}} +b_{e_{d_j} } = \phi {}_j ,  
\end{equation}
where $e_1 , e_2 , \ldots , e_{d_j}$ are arcs emanating from $j$, and $d_j = \deg j$.  
By condition 3, we have 
\begin{equation}
\sum^{d_j }_{r=1} (a_{e_r} -b_{e_r } )={\bf i} \lambda {}_j \phi {}_j 
= \frac{{\bf i} \lambda {}_j }{d_j } \sum^{d_j }_{r=1} (a_{e_r} +b_{e_r } ) .  
\end{equation}

Thus, 
\begin{equation}
\sum^{d_j }_{r=1} b_{e_r } = \frac{1- {\bf i} \lambda {}_j /k d_j }{1+ {\bf i} \lambda {}_j /k d_j } 
\sum^{d_j }_{r=1} a_{e_r} .  
\end{equation}
By (4), for $1 \leq p \leq d_j $, we have 
\[
b_{e_p} = \phi {}_j - a_{e_p} = \frac{1}{d_j} \sum^{d_j }_{r=1} (a_{e_r} +b_{e_r } )- a_{e_p} . 
\]
By (6), 
\begin{equation}
b_{e_p} = \sum^{d_j }_{r=1} \left( \frac{2{\bf i}k}{{\bf i}kd_j - \lambda {}_j } - \delta {}_{e_r e_p} \right) a_{e_r} ,  
\end{equation} 
where $\delta {}_{e_r e_p }$ is the Kronecker delta. 
By (3) and (7), we have 
\begin{equation}
a_{e_p} = b_{e^{-1}_p} e^{{\bf i} L_{e_P} (k+ A_{e_p} )} 
= \sum^{d_l }_{r=1} ( \frac{2{\bf i}k}{{\bf i}kd_l - \lambda {}_l } - \delta {}_{f_r e^{-1}_p} ) 
e^{{\bf i} L_{e_p} (k+ A_{e_p} )} a_{f_r} , 
\end{equation}
where $f_1 , \ldots , f_{d_l}$ are arcs emanating from $l$. 

Now, we introduce the Gnutzmann-Smilansky type of the bond scattering matrix 
of a quantum graph. 
Let 
\[
c_{e^{-1}} =a_e \  for \  each \  e \in D(G) . 
\]
Then we have 
\[
c_{e^{-1}_p} = \sum^{d_l }_{r=1} \left( \frac{2{\bf i}k}{{\bf i}kd_l - \lambda {}_l } - \delta {}_{f^{-1}_r e_p} \right) 
e^{{\bf i} L_{e^{-1}_p} (k- A_{e^{-1}_p} )} c_{f^{-1}_r} . 
\]
Thus, for each arc $e$ with $o(e)=l$, 
\begin{equation}
c_e = \sum_{t(f)=l} \sigma {}^{(l)}_{ef} (k) e^{{\bf i} L_{e} (k- A_e )} c_f , 
\end{equation}
where 
\[
\sigma {}^{(l)}_{ef} (k)= \frac{2{\bf i}k}{{\bf i}kd_l - \lambda {}_l } 
- \delta {}_{e^{-1} f} .  
\]
The {\em vertex scattering matrix} ${\bf S} (k)=( S_{ef} (k))_{e,f \in D(G)} $ 
of $G$ is defined by 
\[
S_{ef} (k)=\left\{
\begin{array}{ll}
\sigma^{(t(f))}_{ef} (k) & \mbox{if $t(f)=o(e)$, } \\
0 & \mbox{otherwise. }
\end{array}
\right. 
\] 

Next, the {\em bond propagation matrix} ${\bf T} (k)=( T_{ef} (k))_{e,f \in D(G)} $ 
of $G$ is defined by 
\[
T_{ef} (k)=\left\{
\begin{array}{ll}
\exp{({\bf i}L_{e} (k - A_{e} ))} & \mbox{if $e=f$, } \\
0 & \mbox{otherwise. }
\end{array}
\right. 
\] 
Then we define the Gnutzmann-Smilansky type of the {\em bond scattering matrix} 
${\bf U}_{GS} (k)= {\bf U}_{GS} (G,k)$ by 
\begin{equation}
{\bf U}_{GS} (k)= {\bf T} (k) {\bf S} (k). 
\end{equation}

By (9), we have 
\begin{equation}
{\bf U}_{GS} (k) {\bf c} = {\bf c} , 
\end{equation}
where ${\bf c} = {}^t (c_1 , c_2 , \ldots , c_{2m} )$. 
Then (9) holds if and only if 
\[
\det ( {\bf I}_{2m} - {\bf U}_{GS} (k))=0 . 
\]

Now, we introduce another type of the bond scattering matrix 
of a quantum graph. 
By (6), 
\begin{equation}
\sum^{d_j }_{r=1} a_{e_r } = \frac{1+ {\bf i} \lambda {}_j /k d_j }{1- {\bf i} \lambda {}_j /k d_j } 
\sum^{d_j }_{r=1} b_{e_r} .  
\end{equation}
By (4), for $1 \leq p \leq d_j $, we have 
\[
a_{e_p} = \phi {}_j - b_{e_p} = \frac{1}{d_j} \sum^{d_j }_{r=1} (a_{e_r} +b_{e_r } )- b_{e_p} . 
\]
By (12), 
\begin{equation}
a_{e_p} = \sum^{d_j }_{r=1} \left( \frac{2{\bf i}k}{{\bf i}kd_j + \lambda {}_j } - \delta {}_{e_r e_p} \right) b_{e_r} . 
\end{equation}
By (3), we have 
\begin{equation}
a_{e_p} = \sum^{d_j }_{r=1} \left( \frac{2{\bf i}k}{{\bf i}kd_j + \lambda {}_j } - \delta {}_{e_r e_p} \right) 
e^{-{\bf i} L_{e_r} (k- A_{e_r} )} a_{e^{-1}_r} .  
\end{equation}

By (8) and (14), we have the following result.

\newtheorem{proposition}{Proposition}
\begin{proposition}
In a quantum graph $G$, for an arc $e=(j,l) \in D(G)$, 
\[
\sum_{o(f)=j} \left( \frac{2{\bf i}k}{{\bf i}kd_j + \lambda {}_j } - \delta {}_{fe} \right) 
e^{-{\bf i} L_f (k- A_f )} a_{f^{-1}} 
= \sum_{o(g)=l} \left( \frac{2{\bf i}k}{{\bf i}kd_l - \lambda {}_l } - \delta {}_{g e^{-1}} \right) 
e^{{\bf i} L_{e} (k+ A_{e} )} a_{g} . 
\] 
\end{proposition}

On the other hand, for an arc $e$ such that $o(e)=j$, (13) is changed into 
\begin{equation}
\begin{array}{rcl}
a_{e} & = & \sum_{t(f)=j} \left( \frac{2{\bf i}k}{{\bf i}kd_j + \lambda {}_j } - \delta {}_{f e^{-1}} \right) 
e^{-{\bf i} L_{f} (k+ A_{f} )} a_{f} \\ 
\  &   &                \\ 
\  & = & \sum_{t(f)=j} \sigma {}^{(j)}_{ef} (-k) e^{-{\bf i} L_{f} (k+ A_f )} a_f ,  
\end{array} 
\end{equation}
where 
\[
\sigma {}^{(j)}_{ef} (-k)= \frac{2{\bf i}k}{{\bf i}k d_j + \lambda {}_j } 
- \delta {}_{e^{-1} f}  . 
\]
The $(e,f)$-array of the vertex scattering matrix  
${\bf S} (-k)=( S_{ef} (-k))_{e,f \in D(G)} $ of $G$ is given by 
\[
S_{ef} (-k)=\left\{
\begin{array}{ll}
\sigma^{(t(f))}_{ef} (-k) & \mbox{if $t(f)=o(e)$, } \\
0 & \mbox{otherwise. }
\end{array}
\right. 
\] 
Furthermore, the $(e,f)$-array of the bond propagation matrix 
${\bf T} (-k)=( T_{ef} (-k))_{e,f \in D(G)} $ of $G$ is given by 
\[
T_{ef} (-k)=\left\{
\begin{array}{ll}
\exp{(-{\bf i}L_{e} (k+ A_{e} ))} & \mbox{if $e=f$, } \\
0 & \mbox{otherwise. }
\end{array}
\right. 
\] 
Then we define another type of the {\em bond scattering matrix} 
${\bf U}_{HKSS} (k)= {\bf U}_{HKSS} (G,k)$ by 
\begin{equation}
{\bf U}_{HKSS} (k)= {\bf S} (-k) {\bf T} (-k) . 
\end{equation}

By (15), we have 
\begin{equation}
{\bf U}_{HKSS} (k) {\bf a} = {\bf a} , 
\end{equation}
where ${\bf a} = {}^t (a_1 , a_2 , \ldots , a_{2m} )$. 
Then (14) holds if and only if 
\[
\det ( {\bf I}_{2m} - {\bf U}_{HKSS} (k))=0 . 
\]

Now, we state the relation between the Gnutzmann-Smilansky scattering matrix and 
another scattering matrix of a quantum graph. 

At first, let $j \in V(G)$, and $e_1 , e_2 , \ldots , e_{d_j}$ be 
arcs emanating from $j$. 
Furthermore, let 
\[
{\bf a}_j = {}^t ( a_{ e_1 } , \ldots , a_{e_{d_j}} ), 
{\bf b}_j = {}^t ( b_{ e_1 } , \ldots , b_{e_{d_j}} ), 
x_j = x_j (k)= \frac{2{\bf i}k}{{\bf i}k d_j - \lambda {}_j } . 
\]
%Then we have 
Then (7) implies that
\[
{\bf b}_j =( x_j {\bf J}_{d_j} - {\bf I}_{d_j } ) {\bf a}_j , \  \  \  \  \  \  (*) 
\]
where ${\bf J}_{d_j} $ is the $d_j \times d_j $ matrix with all one. 
%Thus, let 
Thus, putting
\[
{\bf F}_j =x_j {\bf J}_{d_j} - {\bf I}_{d_j } . 
\]
%Then we have 
the above equation is reexpressed by 
\[
{\bf b}_j = {\bf F}_j {\bf a}_j .  
\]
Here 
\[
\det {\bf F}_j =( d_j x_j -1)(-1 )^{d_j -1} \neq 0 
\]
and 
\[
{\bf F}^{-1}_j = x_j (-k) {\bf J}_{d_j} - {\bf I}_{d_j } . 
\]
Let 
\[
{\bf a} = {}^t ( {\bf a}_1 , \ldots , {\bf a}_n ), 
{\bf b} = {}^t ( {\bf b}_1 , \ldots , {\bf b}_n ), 
{\bf F} = {\bf F}_1 \oplus \ldots \oplus {\bf F}_n . 
\]
Then 
we have 
\begin{equation}
{\bf b} = {\bf F} {\bf a}  \  and \  {\bf a} = {\bf F}^{-1} {\bf b} . 
\end{equation}

Next, let the $2m \times 2m$ diagonal matrix ${\bf R} (k)=( R_{ef} (k))$ be given by 
\[
R_{ef} (k)=\left\{
\begin{array}{ll}
e^{{\bf i} L_e (k+ A_e )} & \mbox{if $e=f$, } \\
0 & \mbox{otherwise. }
\end{array}
\right. 
\] 
Since $a_e = b_{e^{-1}} e^{{\bf i} L_e (k+ A_e )} $, we have 
\begin{equation}
{\bf a} = {\bf R} (k) {\bf J}_0 {\bf b} ,  
\end{equation}
where ${\bf J}_0 =( J_{ef})$ is given by 
\[
J_{ef} =\left\{
\begin{array}{ll}
1 & \mbox{if $f=e^{-1} $, } \\
0 & \mbox{otherwise. }
\end{array}
\right. 
\] 
Note that ${\bf J}^{-1}_0 ={\bf J}_0 $.  
By (18) and (19), (8) is rewritten as follows: 
\begin{equation}
{\bf a} = {\bf R} (k) {\bf J}_0 {\bf F} {\bf a} . 
\end{equation}
Furthermore, by (19), 
\[
{\bf b} = {\bf J}_0 {\bf R} (k)^{-1} {\bf a} , 
\]
and so, (14) is also rewritten as follows: 
\begin{equation}
{\bf a} = {\bf F}^{-1} {\bf b} = {\bf F}^{-1} {\bf J}_0 {\bf R} (k)^{-1} {\bf a} . 
\end{equation}
By (20) and (21), 
%Proposition 1 is given as follows: 
we obtain the following equivalent expression to Proposition 1: 
\[ 
{\bf a} = {\bf R} (k) {\bf J}_0 {\bf F} {\bf a} 
= {\bf F}^{-1} {\bf J}_0 {\bf R} (k)^{-1} {\bf a} . 
\] 

By the way, it holds that 
\[
{\bf T} (k)= {\bf J}_0 {\bf R} (k) {\bf J}_0 \  and \  {\bf S} (k)= {\bf F} {\bf J}_0 . 
\]
Thus, 
\begin{equation}
{\bf U}_{GS} (k)={\bf T} (k) {\bf S} (k)
= {\bf J}_0 {\bf R} (k) {\bf J}_0 {\bf F} {\bf J}_0 . 
\end{equation}
Furthermore, we have 
\[
{\bf T} (-k)= {\bf R} (k)^{-1} \  and \  {\bf S} (-k)= {\bf F}^{-1} {\bf J}_0 . 
\]
Thus, 
\begin{equation}
{\bf U}_{HKSS} (k)={\bf S} (-k) {\bf T} (-k)
= {\bf F}^{-1} {\bf J}_0 {\bf R} (k)^{-1} .
\end{equation}
By (22), (23), we obtain the following result.

\begin{proposition}
In a quantum graph $G$, 
\[
{\bf U}_{GS} (k)= {\bf J}_0 {\bf U}^{-1}_{HKSS} (k) {\bf J}_0 . 
\] 
\end{proposition} 

%%%%%%%%%%%%%%%%%%%%%%%%%%%%%%%%%%%%%%
\section{Quantum graph walks}
%%%%%%%%%%%%%%%%%%%%%%%%%%%%%%%%%%%%%%

At first, we state a short review on a discrete-time quantum walk on a graph. 

Let $G$ be a graph with $n$ vertices and $m$ edges. 
For $v \in V(G)$, let $N^+ (v)= \{ e \in D(G) \mid o(e)=v \} $. 
The we consider a quantum walk over $D(G)$. 
For each arc $e=(u,v) \in D(G)$, the {\em pure state} is given by 
$\vec{x}_{e} = \vec{x}_{uv} =|e \rangle =|u,v \rangle \in {\mathbb C}^{2m} $ such that 
$\{ |e \rangle  \mid e \in D(G) \} $ is the normal orthogonal system of 
the $2m$-dimensional Hilbert space ${\mathbb C}^{2m} $. 
${\cal H} = \ell {}^2 (D(G))= {\rm span} \{ |e \rangle \mid e \in D(G) \} $ is called the 
{\em total space} of a quantum walk on $G$. 
Then we have  
\[
{\cal H} = \bigoplus_{v \in V(G)} {\cal H}_v \  and \  
{\cal H}_v \cong {\rm span} \{ |e \rangle \mid e \in N^+ (v) \} . 
\]

Let $(u,v),(w,x) \in D(G)$. 
Then the {\em transition} from $(u,v)$ to $(w,x)$ occurs if $v=w$. 
The {\em state} $ \psi $ of a quantum walk on $G$ is defined by 
\[
\psi = \sum_{e \in D(G)} \alpha {}_e |e \rangle , \  \alpha {}_e \in {\mathbb C} , 
\]
where  $\sum_{e \in D(G)} | \alpha {}_e |^2 =1$. 
Furthermore, the {\em probability} which the walk is at the arc $e$ is given by 
$| \alpha {}_e |^2 $. 

The {\em time evolution} of a quantum walk on $G$ is given by a unitary matrix ${\bf U} $. 
By the definition of the transition, ${\bf U} =( U_{ef})_{e,f \in D(G)} $ is given as follows 
so that ${\bf U} $ is unitary: 
\[
U_{ef} = \left\{
\begin{array}{ll}
{\rm nonzero \ complex \ number} & \mbox{if $t(e)=o(f)$ (or $t(f)=o(e)$), } \\
0 & \mbox{otherwise. }
\end{array}
\right. 
\] 
For an initial state $\psi {}_0 $ with $|| \psi {}_0 ||=1$, the time evolution is 
the iteration $\psi {}_0 \mapsto \psi {}_1  \mapsto \ldots $ of ${\bf U} $ such that 
\[
\psi {}_j = {\bf U}^j \psi {}_0 , \  j \in {\bf N} . 
\]

Now, we explain a quantum walk called {\em coined quantum walks} on a graph $G$. 
Set $V(G)= \{ 1, \ldots , n \} $. 
Then we choose a sequence of unitary operators $\{ {\bf H}_j \} {}_{j \in V(G)} $,  
where ${\bf H}_j $ is a $d_j $-dimensional operator on ${\cal H}_j $. 
Then we present two types of time evolutions ${\bf U}^{(G)} $ and ${\bf U}^{(A)} $ of 
quantum walks, respectively: 
\[
{\bf U}^{(G)} = {\bf H} {\bf J}_0 ; {\bf U}^{(A)} = {\bf J}_0 {\bf H} , 
\]
where ${\bf H} = \bigoplus_{j \in V(G)} {\bf H}_j $. 
${\bf U}^{(G)} $ and ${\bf U}^{(A)} $ are called {\em Gudder type} and {\em Ambainis type}, 
respectively. 
The elements of ${\bf U}^{(G)} $(or ${\bf U}^{(A)} $) is nonzero if $t(f)=o(e)$ (or $t(e)=o(f)$). 
The first type determined by ${\bf U}^{(G)}$ is a generalization of Gudder [12] (1988) of $d$-dimensional lattice case. 
The second one ${\bf U}^{(A)}$ is motivated by the most popular time evolution for the study of QWs by Ambainis et al [2] 
(2001). 

Next, we treat a quantum graph walk. 
Let $G$ be a connected graph $n$ vertices $1, \ldots , n$, and $m$ edges, and 
let $L: D(G) \longrightarrow {\mathbb R}^+ $ and $A: D(G) \longrightarrow {\mathbb R}$ be 
the length and the magnetic flux of arcs of $G$, respectively. 
Let $ \lambda : V(G) \longrightarrow {\mathbb C} $ be the parameters in the boundary condition III. 
The {\em quantum graph walk} with parameters $(L,A, \lambda )$ is defined as a quantum walk on $G$ 
by the Ambainis type time evolution $\tilde{{\bf U}} $ with the flip flop ${\bf J}_0 $ and the 
following local quantum coin ${\bf H}_j =( (H_j )_{ef} )_{e,f \in N^+ (j)}$ at a vertex $j \in V(G)$: 
\[
(H_j )_{ef} = \left( \frac{2 {\bf i} k}{d_j {\bf i} k- \lambda {}_j } - \delta {}_{ef} \right) 
e^{{\bf i} L_e (k- A_e)} , \  e,f \in N^+ (j) . 
\]
Note that 
\begin{equation}
\tilde{{\bf U}} = {\bf J}_0 {\bf H} , \  {\bf H} = {\bf H}_1 \oplus \cdots \oplus {\bf H}_n . 
\end{equation} 
For brevity, this quantum graph walk is denoted by $\tilde{{\bf U}} $. 
%But,
By the way, 
%we have 
the quantum coin is reexpressed by  
\[
{\bf H} = {\bf T} (k) {\bf F} . 
\]
Furthermore, recall that 
\[
{\bf T} (k)= {\bf J}_0 {\bf R} (k) {\bf J}_0 ,  
\]
%and so, 
Using these relation implies
\begin{equation}
{\bf H} = {\bf J}_0 {\bf R} (k) {\bf J}_0 {\bf F} .  
\end{equation}  
By (24) and (25), we have 
\begin{equation}
\tilde{{\bf U}} = {\bf J}_0 {\bf T} (k) {\bf F} ={\bf R} (k) {\bf J}_0 {\bf F} . 
\end{equation} 
By (20), (8) is rewritten as follows: 
\begin{equation}
{\bf a} = \tilde{{\bf U}} {\bf a} . 
\end{equation} 

Next, we can interpret two scattering matrices ${\bf U}_{GS} (k)$, and ${\bf U}_{HKSS} (k)$ which have discussed in 
the previous section as two kinds of quantum graph walks in the following sence. 
%Next, we introduce two quantum graph walks with parameters $(L,A, \lambda )$. 
%One is a quantum graph walk of $G$ with the time evolution ${\bf U}_{GS} (k)$. 
%Another is a quantum graph walk of $G$ with the time evolution ${\bf U}_{HKSS} (k)$. 
By (22) and (26), we have 
\begin{equation}
{\bf U}_{GS} (k)= {\bf J}_0 {\bf R} (k) {\bf J}_0 {\bf F} {\bf J}_0 = {\bf J}_0 \tilde{{\bf U}} {\bf J}_0 . 
\end{equation} 
By (23) and (26), we have 
\begin{equation}
{\bf U}_{HKSS} (k)= {\bf F}^{-1} {\bf J}_0 {\bf R} (k)^{-1} = \tilde{{\bf U}}^{-1} . 
\end{equation} 
By the forms of ${\bf U}_{GS} (k)$ and ${\bf U}_{HKSS} (k)$, ${\bf U}_{GS} (k)$ and ${\bf U}_{HKSS} (k)$ are Gudder type 
quantum graph walks. 
Furthermore, we introduce the third quantum graph walk of $G$ with the following time evolution: 
\begin{equation}
{\bf U}^{\prime } = {\bf J}_0 {\bf H}^{-1} . 
\end{equation} 
This is an Ambainis type quantum graph walk. 

%For four quantum graph walk, the following result holds. 
As a consequence, the following result in relation to the quantum graph and corresponding four kinds of quantum graph walks holds. 

\newtheorem{theorem}{Theorem}
\begin{theorem}
In the quantum graph $G$ with parameters $(L,A, \lambda )$, 
the following statements are equivalent: 
\begin{enumerate}
\item The Schr\"{o}dinger equation (1) with the boundary conditions I. II. III 
has a nontrivial solution $\{ \Psi {}_e \} {}_{e \in D(G)} $; 
\item The time evolution $\tilde{{\bf U}} $ of the quantum graph walk has the eigenvalue 1. 
\item The time evolution ${\bf U}_{GS} (k) $ of the quantum graph walk has the eigenvalue 1.
\item The time evolution ${\bf U}_{HKSS} (k) $ of the quantum graph walk has the eigenvalue 1. 
\item The time evolution ${\bf U}^{\prime } $ of the quantum graph walk has the eigenvalue 1.
\end{enumerate} 
\end{theorem}

{\bf Proof}. (1) $\Leftrightarrow$ (2): By Theorem 5 of [16].  

(2) $\Leftrightarrow$ (3):  Since ${\bf J}_0 {\bf a}= {\bf c}$, (27) and (28) implies 
that 
\[
\begin{array}{rcl} 
{\bf a} = \tilde{{\bf U}} {\bf a} & \Leftrightarrow & 
{\bf J}_0 {\bf a} = {\bf J}_0 \tilde{{\bf U}} {\bf J}_0 {\bf J}_0 {\bf a} \\
\  &   &                \\ 
\  & \Leftrightarrow & {\bf c} = {\bf U}_{GS} (k) {\bf c} . 
\end{array}
\]

(2) $\Leftrightarrow$ (4):  By (29), 
\[
{\bf a} = \tilde{{\bf U}} {\bf a} \Leftrightarrow 
{\bf a} = \tilde{{\bf U}}^{-1} {\bf a} = {\bf U}_{HKSS} (k) {\bf a} . 
\] 

(2) $\Leftrightarrow$ (5):  By (30), 
\[
\begin{array}{rcl} 
{\bf a} = \tilde{{\bf U}} {\bf a} & \Leftrightarrow & 
{\bf a} = \tilde{{\bf U}}^{-1} {\bf a} = {\bf H}^{-1} {\bf J}_0 {\bf a} \\
\  &   &                \\ 
\  & \Leftrightarrow & {\bf J}_0 {\bf a} = {\bf J}_0 {\bf H}^{-1} {\bf J}_0 {\bf a} \\
\  &   &                \\ 
\  & \Leftrightarrow & {\bf c} = {\bf U}^{\prime } {\bf c} . 
\end{array}
\]
\begin{flushright}$\Box$\end{flushright} 

Note that if ${\bf a} = \Phi $ is the eigenvector for the eigenvalue 1 of $\tilde{{\bf U}} $, 
then ${\bf a} = \Phi $ is the eigenvector for the eigenvalue 1 of ${\bf U}_{HKSS} (k)$, 
and ${\bf J}_0 {\bf a} $ is the eigenvector for the eigenvalue 1 of ${\bf U}_{GS} (k)$ 
and ${\bf U}^{\prime } $. 

%%%%%%%%%%%%%%%%V'µ'­"ü'ê'Ä'Ý'½'Æ'±'ë 'S''Ìwalk'ÌŠÖŒW«%%%%%%%%%%%%%%%%%%%%%%%%%%%%
Finally, we mention a relationship between four quantum graph walks from view point of spatial and temporal duality relation. 
See also Fig.1. 
The quantum graph walks $\tilde{{\bf U}}$ and ${\bf U}_{GS}(k)$ are in a time reversal relation in that 
$\tilde{{\bf U}}^{-1}={\bf U}_{GS}(k)$. We can see also the same time reversal relation between ${\bf U}_{HKSS}$ and ${\bf U}'$. 
On the other hand, $\tilde{{\bf U}}$ and ${\bf U}_{HKSS}(k)$ are in a spatial reversal relation in that 
${\bf J}_0\tilde{{\bf U}}{\bf J}_0={\bf U}_{GS}(k)$, that is, the total space of $\tilde{{\bf U}}$ is descreibed by 
$\bigoplus_{v\in V(G)}\mathrm{span}\{ |e\rangle | e\in N^+(v) \}$, 
while the total space of $\tilde{{\bf U}}$ is descreibed by 
$\bigoplus_{v\in V(G)}\mathrm{span}\{ |e\rangle | e\in N^-(v) \}$,
where $N^-(v)=\{e\in D(G)|t(e)=v\}$. 
We can see also the same spatial reversal relation between ${\bf U}_{GS}$ and ${\bf U}'$. 

\begin{figure}
\begin{center}
\includegraphics[width=5cm]{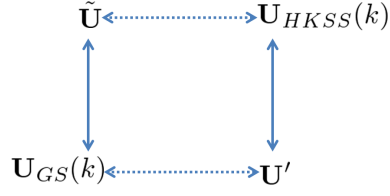}
\caption{ {\scriptsize Spatial and temporal duality relationship of four quantum graph walks: 
The solid lines (vertical lines) depict the spatial reversal relationship in that 
${\bf J}_0 \tilde{{\bf U}} {\bf J}_0 ={\bf U}_{GS}(k)$ and ${\bf J}_0 {\bf U}_{HKSS}(k) {\bf J}_0 ={\bf U}'$. 
The dotted lines (horizontal lines) express the temporal reversal relationship in that 
$\tilde{{\bf U}}^{-1} ={\bf U}_{HKSS}(k)$ and ${\bf U}_{GS}(k)^{-1} ={\bf U}'$. }
}
\end{center}
\end{figure}
%%%%%%%%%%%%%%%%%%%%%%%%%%%%%%%%%%%%%%%%%%%%

%%%%%%%%%%%%%%%%%%%%%%%%%%%%%%%%%%%%%%%%%%%%
%%%%%%%%%%%%%%%%%%%%%%%%%%%%%%%%%%%%%%%%%%%%
\section{The characteristic polynomial of a scattering matrix of a quantum graph}

Let $G$ be a connected graph with $n$ vertices and $m$ unoriented edges. 
Set $V(G)= \{ 1,2, \ldots, n \}$ and $D(G)= \{ e_1 , e^{-1}_1 , \ldots , e_m , e^{-1}_m \} $. 
Furthermore, for $j \in V(G)$ and $e \in D(G)$, let 
\[
x_j = \frac{2{\bf i}k}{{\bf i}k d_j - \lambda {}_j } \  and \  
t_e = \exp{({\bf i}L_{e} (k - A_{e} ))} . 
\]
Furthermore, set 
\[
\sigma^{(t(f))}_{ef} = \sigma^{(t(f))}_{ef} (k) , 
{\bf U} ={\bf U} (G)={\bf U}_{GS} (k), {\bf T} = {\bf T} (k) , {\bf S} = {\bf S} (k) . 
\]
 
Let an $n \times n$ matrix 
$\tilde{{\bf A}} =\tilde{{\bf A}} ( \sigma {}^{2})=\tilde{{\bf A}} (G, \sigma {}^{2})=( \tilde{a}_{uv} )$ 
be defined by
\[
\tilde{a}_{uv} =\left\{
\begin{array}{ll}
\frac{x_v t_e }{\sigma {}^2 - t_e t_{e^{-1}} } & \mbox{if $e=(u,v) \in D(G)$, } \\
0 & \mbox{otherwise.}
\end{array}
\right. 
\  \  \  \  \  \  \  (**)  
\] 
Let an $n \times n$ matrix 
$\overline{{\bf A}} = \overline{{\bf A}} ( \sigma {}^{2})=\overline{{\bf A}} (G, \sigma{}^{2})=( \overline{a}_{uv} )$ 
be defined by
\[
\overline{a}_{uv} =\left\{
\begin{array}{ll}
\frac{x_u t_e }{\sigma {}^2 - t_e t_{e^{-1}}} & \mbox{if $e=(u,v) \in D(G)$, } \\
0 & \mbox{otherwise.}
\end{array}
\right. 
\  \  \  \  \  \  \  (***)
\] 
Furthermore, let an $n \times n$ diagonal matrix 
$\overline{{\bf D}} =\overline{{\bf D}} ( \sigma {}^{2})=\overline{{\bf D}} (G, \sigma {}^{2} )=( d_{uv} )$ be defined by
\[
d_{uv} =\left\{
\begin{array}{ll}
\sum_{o(e)=u} \frac{ x_u t_e t_{e^{-1}} }{\sigma {}^2 - t_e t_{e^{-1} }} & \mbox{if $u=v$, } \\
0 & \mbox{otherwise.}
\end{array}
\right.
\] 
Note that $t_e t_{e^{-1} } =1- e^{2ikL_e } , e \in D(G)$.

\begin{theorem}
Let $G$ be a connected graph with $n$ vertices and $m$ unoriented edges. 
Then 
\[
\det ( \sigma {\bf I}_{2m} - {\bf U} )  
= \det ( {\bf I}_{n} - \sigma \tilde{{\bf A}} + \overline{{\bf D}} ) 
\prod^m_{j=1} (\sigma {}^2 - e^{2{\bf i}kL_{e_j} } ) 
= \det ( {\bf I}_{n} - \sigma \overline{{\bf A}} + \overline{{\bf D}} ) 
\prod^m_{j=1} (\sigma {}^2 - e^{2{\bf i}kL_{e_j} } ) . 
\] 
\end{theorem}

{\bf Proof}.  The argument is an analogue of the method of Watanabe and Fukumizu [40].  

Let $D(G)= \{ e_1, \ldots , e_m , e_{m+1},$ 
$\ldots , e_{2m} \} $ such that 
$ e_{m+i} = e^{-1}_i (1 \leq i \leq m)$. 
Furthermore, arrange arcs of $G$ as follows: 
\[
e_1 , e^{-1}_1 , \ldots , e_m , e^{-1}_m . 
\]  
Note that the $(e,f)$-array $( {\bf U} )_{ef} $ of ${\bf U} $ is given by 
\[ 
( {\bf U} )_{ef} 
=\left\{
\begin{array}{ll}
t_e ( x_{o(e) } - \delta {}_{e^{-1} f} ) & \mbox{if $t(f)=o(e)$, } \\
0 & \mbox{otherwise.}
\end{array}
\right.
\] 

Let $2m \times 2m$ matrices ${\bf B} =( {\bf B}_{ef} )_{e,f \in D(G)} $ and 
${\bf J}_0 =( {\bf J}_{ef} )_{e,f \in D(G)} $ be defined as follows: 
\[
{\bf B}_{ef} =\left\{
\begin{array}{ll}
x_{t(e)} & \mbox{if $t(e)=o(f)$, } \\
0 & \mbox{otherwise, }
\end{array}
\right.
\  
{\bf J}_{ef} =\left\{
\begin{array}{ll}
1 & \mbox{if $f= e^{-1} $, } \\
0 & \mbox{otherwise.}
\end{array}
\right.
\]
Note that ${}^t {\bf J}_0 ={\bf J}_0 $.  

Now  
\begin{equation}
{}^t  {\bf S} = {\bf B} -{\bf J}_0 .  
\end{equation}
 
Let ${\bf K} =( {\bf K}_{ev} )$ ${}_{e \in D(G); v \in V(G)} $ be 
the $2m \times n$ matrix defined 
as follows: 
\[
{\bf K}_{ev} :=\left\{
\begin{array}{ll}
1 & \mbox{if $o(e)=v$, } \\
0 & \mbox{otherwise. } 
\end{array}
\right.
\]
Furthermore, we define a $2m \times n$ matrix  
${\bf L} =( {\bf L}_{ev} ) {}_{e \in D(G); v \in V(G)} $ as follows: 
\[
{\bf L}_{ev} :=\left\{
\begin{array}{ll}
1 & \mbox{if $t(e)=v$, } \\
0 & \mbox{otherwise. } 
\end{array}
\right.
\]
Then we have  
\begin{equation}
{\bf L} {\bf X} {}^t {\bf K} = {\bf B} ,    
\end{equation} 
where 
\[
{\bf X} =
\left[
\begin{array}{ccc}
x_{1} & & 0 \\
  & \ddots &  \\
0 &  & x_{n} 
\end{array}
\right],
\]

Next, By (31) and (32), we have  
\[
\begin{array}{rcl}
\  &   & \det ( {\bf I}_{2m} -s {\bf U} )= \det ( {\bf I}_{2m} -s {\bf T} {\bf S}) 
=\det ( {\bf I}_{2m} -s {\bf T} ( {}^t {\bf B} - {\bf J}_0 )) \\ 
\  &   &                \\ 
\  & = & \det ( {\bf I}_{2m} -s {\bf T} ( {\bf K} {\bf X} {}^t {\bf L} - {\bf J}_{0} )) 
= \det ( {\bf I}_{2m} +s {\bf T} {\bf J}_{0} -s {\bf T} {\bf K} {\bf X} {}^t {\bf L} ) \\
\  &   &                \\ 
\  & = & \det ( {\bf I}_{2m} -s {\bf T} {\bf K} {\bf X} {}^t {\bf L} 
( {\bf I}_{2m} +s {\bf T} {\bf J}_0 )^{-1} ) 
\det ( {\bf I}_{2m} +s {\bf T} {\bf J}_0 ) . 
\end{array}
\]

If ${\bf A}$ and ${\bf B}$ are a $m \times n $ and $n \times m$ 
matrices, respectively, then we have 
\begin{equation}
\det ( {\bf I}_{m} - {\bf A} {\bf B} )= 
\det ( {\bf I}_n - {\bf B} {\bf A} ) . 
\end{equation}
Thus, we have 
\begin{equation} 
\det ( {\bf I}_{2m} -s {\bf U} )
= \det ( {\bf I}_{2m} -s {\bf X} {}^t {\bf L} 
( {\bf I}_{2m} +s {\bf T} {\bf J}_0 )^{-1} {\bf T} {\bf K} ) 
\det ( {\bf I}_{2m} +s {\bf T} {\bf J}_0) . 
\end{equation} 
Furthermore, 
\begin{equation}
\det ( {\bf I}_{2m} -s {\bf U} )
= \det ( {\bf I}_{2m} -s {}^t {\bf L} ( {\bf I}_{2m} +s {\bf T} {\bf J}_0 )^{-1} 
{\bf T} {\bf K} {\bf X} )
\det ( {\bf I}_{2m} +s {\bf T} {\bf J}_0) . 
\end{equation} 

Next, we have 
\[
{\bf I}_{2m} +s {\bf T} {\bf J}_0 =
\left[ 
\begin{array}{ccccc}
1 & st_{e_1 } &   &   & {\bf 0} \\
st_{e^{-1}_1 } & 1 &   &  \\ 
   &    & \ddots &   &   \\
   &    &   & 1 & st_{e_m} \\ 
{\bf 0} &    &   & st_{e^{-1}_m } & 1 
\end{array} 
\right] 
,  
\]
and so, 
\[
\det ( {\bf I}_{2m} +s {\bf T} {\bf J}_0 ) 
= \prod^{m}_{j=1} (1- t_{e_j  } t_{e^{-1}_j } s^2 ) . 
\]
Furthermore, we have 
\[
( {\bf I}_{2m} +s {\bf T} {\bf J}_0 )^{-1} 
= 
\left[ 
\begin{array}{ccc}
1/y_1  & -s t_{e_1 } /y_1 & {\bf 0} \\ 
-s t_{e^{-1}_1 } /y_1 & 1/y_1 \\
{\bf 0}  &   &  \ddots 
\end{array} 
\right] 
,  
\]
where $y_j =1- t_{e_j} t_{e^{-1}_j } s^2 \  (1 \leq j \leq m)$.  

For an arc $(u,v) \in D(G)$, 
\[
( {\bf X} {}^t {\bf L} ( {\bf I}_{2m} +s {\bf T} {\bf J}_0 )^{-1} 
{\bf T} {\bf K} )_{uv} 
= x_u t_{(v,u)} /(1- t_{(u,v)} t_{(v,u)} s^2 ) . 
\]
Furthermore, if $u=v$, then 
\[
( {\bf X} {}^t {\bf L} ( {\bf I}_{2m} +s {\bf T} {\bf J}_0 )^{-1} 
{\bf T} {\bf K} )_{uu} 
=- \sum_{t(e)=u} \frac{ x_u t_e t_{e^{-1}} s}{1- t_e t_{e^{-1} } s^2 } . 
\]
Then we have 
\[
{\bf X} {}^t {\bf L} ( {\bf I}_{2m} +s {\bf T} {\bf J}_0 )^{-1} 
{\bf T} {\bf K} =1/s^2 \  {}^t \tilde{{\bf A}} (1/s^2 )-1/s \overline{{\bf D}} (1/s^2 ) . 
\]
Therefore, by (34), it follows that 
\[
\begin{array}{rcl}
\  &   & \det ( {\bf I}_{2m} -s {\bf U} ) \\  
\  &   &                \\ 
\  & = & \det ( {\bf I}_n -s {\bf X} {}^t {\bf L} ( {\bf I}_{2m} 
+s {\bf T} {\bf J}_0 )^{-1} {\bf T} {\bf K} ) \prod^{m}_{j=1} (1- t_{e_j  } t_{e^{-1}_j } s^2 ) \\
\  &   &                \\ 
\  & = & \det ( {\bf I}_n -1/s \  {}^t \tilde{{\bf A}} (1/s^2 )+ \overline{{\bf D}} (1/s^2 )) 
\prod^{m}_{j=1} (1- t_{e_j  } t_{e^{-1}_j } s^2 ) \\
\  &   &                \\ 
\  & = & \det ( {\bf I}_n -1/s  \tilde{{\bf A}} (1/s^2 )+ \overline{{\bf D}} (1/s^2 )) 
\prod^{m}_{j=1} (1- t_{e_j  } t_{e^{-1}_j } s^2 ) . 
\end{array}
\]

Next, for an arc $(u,v) \in D(G)$, 
\[
( {}^t {\bf L} ( {\bf I}_{2m} +s {\bf T} {\bf J}_0 )^{-1} 
{\bf T} {\bf K} {\bf X} )_{uv} 
= x_v t_{(v,u)} /(1- t_{(u,v)} t_{(v,u)} s^2 ) . 
\]
Furthermore, if $u=v$, then 
\[
( {\bf X} {}^t {\bf L} ( {\bf I}_{2m} +s {\bf T} {\bf J}_0 )^{-1} 
{\bf T} {\bf K} {\bf X} )_{uu} 
=- \sum_{o(e)=u} \frac{ x_u t_e t_{e^{-1}} s}{1- t_e t_{e^{-1} } s^2 } .  
\]
Then we have 
\[
{}^t {\bf L} ( {\bf I}_{2m} +s {\bf T} {\bf J}_0 )^{-1} {\bf T} {\bf K} {\bf X}  
=1/s^2 \  {}^t \overline{{\bf A}} (1/s^2 )-1/s \overline{{\bf D}} (1/s^2 ) . 
\]
Therefore, by (35), it follows that 
\[
\begin{array}{rcl}
\  &   & \det ( {\bf I}_{2m} -s {\bf U} ) \\  
\  &   &                \\ 
\  & = & \det ( {\bf I}_{n} -s {}^t {\bf L} ( {\bf I}_{2m} +s {\bf T} {\bf J}_0 )^{-1} {\bf T} {\bf K} {\bf X} ) 
\prod^{m}_{j=1} (1- t_{e_j  } t_{e^{-1}_j } s^2 ) \\
\  &   &                \\ 
\  & = & \det ( {\bf I}_{n} -1/s \   {}^t \overline{{\bf A}} (1/s^2 )+ \overline{{\bf D}} (1/s^2 )) 
\prod^{m}_{j=1} (1- t_{e_j  } t_{e^{-1}_j } s^2 ) \\
\  &   &                \\ 
\  & = & \det ( {\bf I}_{n} -1/s \overline{{\bf A}} (1/s^2 )+ \overline{{\bf D}} (1/s^2 )) 
\prod^{m}_{j=1} (1- t_{e_j  } t_{e^{-1}_j } s^2 ) . 
\end{array}
\]

Now, let $s=1/ \sigma $. Then we get 
\[ 
\det \left( {\bf I}_{2m} - \frac{1}{ \sigma } {\bf U} \right) = 
\prod^{m}_{j=1} (1- t_{e_j  } t_{e^{-1}_j } \frac{1}{ \sigma {}^2 } )  
\det \left( {\bf I}_{n} - \sigma \tilde{{\bf A}} ( \sigma {}^2 )+ \overline{{\bf D}} (\sigma {}^2 ) \right) . 
\]
Thus, 
\[
\det ( \sigma {\bf I}_{2m} - {\bf U} )= 
\prod^{m}_{j=1} (\sigma {}^2 - t_{e_j  } t_{e^{-1}_j } ) 
\det \left( {\bf I}_{n} - \sigma \tilde{{\bf A}} ( \sigma {}^2 )+ \overline{{\bf D}} (\sigma {}^2 ) \right) . 
\]
Furthermore, we have  
\[ 
\det \left( {\bf I}_{2m} - \frac{1}{ \sigma } {\bf U} \right) = 
\prod^{m}_{j=1} (1- t_{e_j  } t_{e^{-1}_j } \frac{1}{ \sigma {}^2 } )  
\det \left( {\bf I}_{n} - \sigma \overline{{\bf A}} ( \sigma {}^2 )+ \overline{{\bf D}} (\sigma {}^2 ) \right) . 
\]
Thus, 
\[
\det ( \sigma {\bf I}_{2m} - {\bf U} )= 
\prod^{m}_{j=1} (\sigma {}^2 - t_{e_j  } t_{e^{-1}_j } ) 
\det \left( {\bf I}_{n} -\sigma \overline{{\bf A}} ( \sigma {}^2 )+ \overline{{\bf D}} (\sigma {}^2 ) \right) . 
\]
\begin{flushright}$\Box$\end{flushright}
%%%%%%%%%%%%%%%%%%%%%%%%%%%%%%%%%%%%%%%%%%%%
%%%%%%%%%%%%%%%%%%%%%%%%%%%%%%%%%%%%%%%%%%%%

%%%%%%%%%%%%%%%%%%%%%%%%%%%%%%%%%%%%%%%%%%%%
%%%%%%%%%%%%%%%%%%%%%%%%%%%%%%%%%%%%%%%%%%%%
\section{The characteristic polynomial of a scattering matrix of a regular covering of a graph}

Let $G$ be a connected graph, and 
let $N(v)= \{ w \in V(G) \mid (v,w) \in D(G) \} $ denote the 
neighbourhood  of a vertex $v$ in $G$. 
A graph $H$ is a {\em covering} of $G$ 
with projection $ \pi : H \longrightarrow G $ if there is a surjection
$ \pi : V(H) \longrightarrow V(G)$ such that
$ \pi {\mid}_{N(v')} : N(v') \longrightarrow N(v)$ is a bijection 
for all vertices $v \in V(G)$ and $v' \in {\pi}^{-1} (v) $.
When a finite group $\Pi$ acts on a graph $G$, 
the {\em quotient graph} $G/ \Pi$ is a graph 
whose vertices are the $\Pi$-orbits on $V(G)$, 
with two vertices being adjacent in $G/ \Pi$ if and only if some two 
of their representatives are adjacent in $G$.
A covering $ \pi : H \longrightarrow G$ is {\em regular} 
if there is a subgroup {\it B} of the automorphism group $Aut \  H$ 
of $H$ acting freely on $H$ such that the quotient graph $H/ {\it B} $ 
is isomorphic to $G$.

Let $G$ be a graph and $ \Gamma $ a finite group.
Then a mapping $ \alpha : D(G) \longrightarrow \Gamma $
is an {\em ordinary voltage} {\em assignment}
if $ \alpha (v,u)= \alpha (u,v)^{-1} $ for each $(u,v) \in D(G)$.
The pair $(G, \alpha )$ is an {\em ordinary voltage graph}.
The {\em derived graph} $G^{ \alpha } $ of the ordinary
voltage graph $(G, \alpha )$ is defined as follows:
$V(G^{ \alpha } )=V(G) \times \Gamma $ and $((u,h),(v,k)) \in 
D(G^{ \alpha })$ if and only if $(u,v) \in D(G)$ and $k=h \alpha (u,v) $. 
The {\em natural projection} 
$ \pi : G^{ \alpha } \longrightarrow G$ is defined by 
$ \pi (u,h)=u$. 
The graph $G^{ \alpha }$ is a {\em derived graph covering} of $G$ 
with voltages in $ \Gamma $ or a {\em $ \Gamma $-covering} of $G$.
The natural projection $ \pi $ commutes with the right 
multiplication action of the $ \alpha (e), e \in D(G)$ and 
the left action of $ \Gamma $ on the fibers: 
$g(u,h)=(u,gh), g \in \Gamma $, which is free and transitive. 
Thus, the $ \Gamma $-covering $G^{ \alpha }$ is a $ \mid \Gamma \mid $-fold
regular covering of $G$ with covering transformation group $ \Gamma $.
Furthermore, every regular covering of a graph $G$ is a 
$ \Gamma $-covering of $G$ for some group $ \Gamma $ (see [10]).
Figure 2 depicts the derived graph of $G=K_3$ with $\Gamma = \mathbb{Z}_2$. 

Let $G$ be a connected graph, $ \Gamma $ be a finite group and 
$ \alpha : D(G) \longrightarrow \Gamma $ be an ordinary voltage assignment. 
In the $\Gamma $-covering $G^{ \alpha } $, set $v_g =(v,g)$ and $e_g =(e,g)$, 
where $v \in V(G), e \in D(G), g\in \Gamma $. 
For $e=(u,v) \in D(G)$, the arc $e_g$ emanates from $u_g$ and 
terminates at $v_{g \alpha (e)}$. 
Note that $ e^{-1}_g =(e^{-1} )_{g \alpha (e)}$.

We consider the Gnutzmann-Smilansky type of the bond scattering matrix of 
the regular covering $G^{\alpha }$ of $G$. 
Let $V(G)= \{ 1 , \ldots , n \} $, 
$D(G)= \{ e_1 , \ldots , e_m , e^{-1}_1 , \ldots , e^{-1}_m \} $ and 
$\Gamma = \{ g_1 =1 , g_2 , \ldots , g_p \} $. 
Let $L: D(G) \longrightarrow {\bf R}^+ $ and $A: D(G) \longrightarrow {\bf R} $ 
be the length and the magnetic flux of arcs of $G$. 
Let the length $\tilde{L}: D(G^{\alpha }) \longrightarrow {\bf R}^+ $ and 
the vector potential $\tilde{A}: D(G^{\alpha }) \longrightarrow {\bf R} $ of arcs of $G^{\alpha }$ 
be given by 
\[
\tilde{L}_{e_g} =L_e \  and \  \tilde{A}_{e_g } = A_e , e \in D(G), g \in \Gamma . 
\]

Let $e=(j,l) \in D(G)$. 
Then we consider the Schr\"{o}dinger equation for the $e_g =(j_g , l_{g \alpha (e)} )$: 
\[
\left(-{\bf i} {\rm \frac{d}{dx} } + \tilde{A}_{e_g } \right)^2 \Psi {}_{e_g } (x)= k^2 \Psi {}_{e_g } (x) 
\]
under the following three conditions: 
\begin{enumerate} 
\item $ \Psi {}_{e_g } (x)= \Psi {}_{e^{-1}_g } (\tilde{L}_{e_g } -x)$; 
\item {\em The continuity}: $\Psi {}_{e_g } (0)= \phi {}_{j_g} $ and 
$\Psi {}_{e_g } (\tilde{L}_{e_g } )= \phi {}_{l_{g \alpha (e)}} $; 
\item {\em The current conservation}: 
\[
\sum_{o(f_g )=j_g } \left(-{\bf i} {\rm \frac{d}{dx} } + \tilde{A}_{f_g } \right) \Psi {}_{f_g } (x) \mid {}_{x=0} 
=-{\bf i} \lambda {}_{j_g} \phi {}_{j_g } , \forall j_g \in V(G^{\alpha } ) ,  
\] 
where $( \phi {}_{1,1} , \ldots , \phi {}_{p, g_m} ) \in {\bf C}^{pm} $. 
\end{enumerate} 

By the definitions of $\tilde{L}$ and $\tilde{A}$, the Schr\"{o}dinger equation for the arc 
$e_g =(j_g , l_{g \alpha (e)} )$ and the three conditions 1,2,3 are reduced to the following system: 
\[
\left(-{\bf i} {\rm \frac{d}{dx} } + A_{e} \right)^2 \Psi {}_{e_g } (x)= k^2 \Psi {}_{e_g } (x) 
\]
and  
\begin{enumerate} 
\item $ \Psi {}_{e_g } (x)= \Psi {}_{e^{-1}_g } (L_e -x)$; 
\item $\Psi {}_{e_g } (0)= \phi {}_{j_g} $ and 
$\Psi {}_{e_g } (L_{e} )= \phi {}_{l_{g \alpha (e)}} $; 
\item  
\[
\sum_{o(f_g )=j_g } \left(-{\bf i} {\rm \frac{d}{dx} } + A_{f} \right) \Psi {}_{f_g } (x) \mid {}_{x=0} 
=-{\bf i} \lambda {}_{j_g} \phi {}_{j_g } , \forall j_g \in V(G^{\alpha } ) .   
\] 
\end{enumerate}

The solution of the Schr\"{o}dinger equation is given by 
\[ 
\Psi {}_{e_g } (x)=( c_{e^{-1}_g } e^{-{\bf i}kx} + b_{e_g } e^{{\bf i}kx}) e^{-{\bf i}A_e x} , {\bf i}= \sqrt{-1} . 
\] 
Similarly to (9), we have 
\[
c_{e_g } = \sum_{t(f_h )=j_g } \sigma {}^{(j_g)}_{e_g f_h} e^{{\bf i} \tilde{L}_{e_g} (k- \tilde{A}_{e_g} )} c_{f_h} , 
\] 
where 
\[
\sigma {}^{(j_g)}_{e_g f_h} = \frac{2{\bf i}ik}{{\bf i}k d_{j_g } - \lambda {}_{j_g } } - \delta {}_{e^{-1}_g f_h} 
= \frac{2{\bf i}k}{{\bf i}k d_j - \lambda {}_{j_g } } - \delta {}_{e^{-1}_g f_h} .  
\]
Then the bond scattering matrix 
${\bf U} ( G^{\alpha } )=( U(e_g , f_h) )_{e_g ,f_h \in D(G^{\alpha } )} $ 
of $ G^{ \alpha } $ is given by 
\begin{equation}
U(e_g , f_h) =\left\{
\begin{array}{ll}
t_{e_g} ( x_{o(e_g ) } - \delta {}_{e^{-1}_g f_h} ) & \mbox{if $t(f_h)=o(e_g)$, } \\
0 & \mbox{otherwise,}
\end{array}
\right.
\end{equation} 
where 
\[
x_{v_g } = \frac{2{\bf i}k}{{\bf i}k d_v - \lambda {}_{v_g} } \  and \  
t_{e_g} = e^{{\bf i} \tilde{L}_{e_g} (k- \tilde{A}_{e_g} )} = e^{{\bf i} L_{e} (k- A_e )} = t_e . 
\]

Now, we assume that 
\[
\lambda {}_{j_g} = \lambda {}_j \  for \  any \  j \in V(G) \  and \  g \in \Gamma . 
\]
Under this assumption, we have 
\[
x_{v_g } = \frac{2{\bf i}k}{{\bf i}k d_v - \lambda {}_{v} } = x_v , \forall v \in V(G), \forall g \in \Gamma . 
\]
Then (36) is reduced to  
\[
U(e_g , f_h) =\left\{
\begin{array}{ll}
t_{e} ( x_{o(e) } - \delta {}_{e^{-1}_g f_h} ) & \mbox{if $t(f_h)=o(e_g)$, } \\
0 & \mbox{otherwise.}
\end{array}
\right.
\]

For $ g \in \Gamma $, let the matrices $\tilde{{\bf A}}_g =\tilde{{\bf A}}_g ( \sigma {}^{2})=( \tilde{a}^{(g)}_{uv} )$ and 
$ \overline{{\bf A} }_g =\overline{{\bf A} }_g ( \sigma {}^{2})=( \overline{a}^{(g)}_{uv} )$ be defined by
\[
\tilde{a}^{(g)}_{uv} =\left\{
\begin{array}{ll}
\frac{x_v t_e }{\sigma {}^2 - t_e t_{e^{-1}} } & \mbox{if $e=(u,v) \in D(G)$ and $ \alpha (e)=g$, } \\
0 & \mbox{otherwise, }
\end{array}
\right.
\]
\[  
\overline{a}^{(g)}_{uv} =\left\{
\begin{array}{ll}
\frac{ x_u t_e }{\sigma {}^2 - t_e t_{e^{-1} }} & \mbox{if $e=(u,v) \in D(G)$ and $ \alpha (e)=g$, } \\
0 & \mbox{otherwise.}
\end{array}
\right.
\]  
Furthermore, let ${\bf U}_g =( U^{(g)} (e,f) )$ be given by 
\[
U^{(g)} (e,f) =\left\{
\begin{array}{ll}
t_e ( x_{o(e) } - \delta {}_{e^{-1} f} ) & \mbox{if $t(f)=o(e)$ and $\alpha (f)=g$, } \\
0 & \mbox{otherwise.}
\end{array}
\right.
\] 
Let ${\bf M}_{1} \oplus \cdots \oplus {\bf M}_{s} $ be the block diagonal sum 
of square matrices ${\bf M}_{1} , \ldots , {\bf M}_{s} $. 
If \( {\bf M}_{1} = {\bf M}_{2} = \cdots = {\bf M}_{s} = {\bf M} \),
then we write 
\( s \circ {\bf M} = {\bf M}_{1} \oplus \cdots \oplus {\bf M}_{s} \).
The {\em Kronecker product} $ {\bf A} \otimes {\bf B} $
of matrices {\bf A} and {\bf B} is considered as the matrix 
{\bf A} having the element $a_{ij}$ replaced by the matrix $a_{ij} {\bf B}$.

\begin{theorem} 
Let $G$ be a connected graph with $n$ vertices and $m$ 
unoriented edges, $ \Gamma $ be a finite group and 
$ \alpha : D(G) \longrightarrow \Gamma  $ be an ordinary voltage assignment. 
Assume that $\tilde{L}_{e_g} =L_e $, $\tilde{A}_{e_g } = A_e $ and 
$\lambda {}_{j_g} = \lambda {}_j $ for  any $e \in D(G), j \in V(G), g \in \Gamma$.  
Set $| \Gamma |=p$. 
Furthermore, let $ {\rho}_{1} =1, {\rho}_{2} , \cdots , {\rho}_{k} $
be the irreducible representations of $ \Gamma $, and 
$f_i$ be the degree of $ {\rho}_{i} $ for each $i$, where $f_1=1$.

If the $ \Gamma $-covering $G^{ \alpha } $ of $G$ is connected, then, 
for the bond scattering matrix of $G {}^{ \alpha } $,   
\[
\det ( \sigma {\bf I}_{2 mp} - {\bf U} ( G^{\alpha } ))= \det( \sigma {\bf I}_{2m} - {\bf U} (G)) 
\prod^{k}_{i=2} \det ( \sigma {\bf I}_{2m f_i } - \sum_{h \in \Gamma } {}^t {\rho}_i (h) \otimes {\bf U} {}_h )^{f_i} 
\]
\[
= \det ( {\bf I}_{n} -\sigma \overline{{\bf A}} + \overline{{\bf D}} ) 
\prod^{k}_{i=2} \det ( {\bf I}_{n f_i } - \sigma \sum_{h \in \Gamma } {\rho}_{i} (h) \otimes \overline{{\bf A}} {}_{h} 
+ {\bf I}_{f_i} \otimes \overline{{\bf D}} )^{f_i} \prod^m_{j=1} (\sigma {}^2 - e^{2{\bf i}kL_{e_j} } )^p  . 
\]
\[
= \det ( {\bf I}_{n} -\sigma \tilde{{\bf A}} + \overline{{\bf D}} ) 
\prod^{k}_{i=2} \det ( {\bf I}_{n f_i } -\sigma \sum_{h \in \Gamma } {\rho}_{i} (h) \otimes \tilde{{\bf A}} {}_{h} 
+ {\bf I}_{f_i} \otimes \overline{{\bf D}} )^{f_i} \prod^m_{j=1} (\sigma {}^2 - e^{2{\bf i}kL_{e_j} } )^p  ,  
\] 
where $D(G)= \{ e_1 , e^{-1}_1 , \ldots , e_m , e^{-1}_m \} $.  
Recall that $\tilde{{\bf A}} $ and $\overline{{\bf A}} $ is defined in (**) and (***), respectively. 
\end{theorem}

{\em Proof }. Let $ \mid \Gamma \mid =p$. 
By Theorem 2, we have  
\[
\det ( \sigma {\bf I}_{2 mp} - {\bf U} ( G^{\alpha } )) 
= \det ( {\bf I}_{np} - \sigma \overline{{\bf A}} (G^{\alpha } , \sigma {}^{2})
+ \overline{{\bf D}} (G^{\alpha } , \sigma {}^{2})) 
\prod^m_{j=1} (\sigma {}^2 - t_{e_j} t_{e^{-1}_j } )^p . 
\]

Let $D(G)= \{ e_1, \ldots , e_m , e_{m+1}, \ldots , e_{2m} \} $ such that 
$e_{m+j} = e^{-1}_j (1 \leq j \leq m)$, and let 
$ \Gamma = \{ 1=g_1, g_2, \ldots ,g_p \} $.
Arrange arcs of $G^{ \alpha } $ in $p$ blocks:
$(e_1,1), \ldots , (e_{2m},1);(e_1,g_2), \ldots , (e_{2m},g_2)$;  
$ \ldots ; (e_1,g_p), \ldots ,(e_{2m},g_p). $
We consider the matrix ${\bf U} ( G^{\alpha } )$
under this order.
For $h \in \Gamma $, the matrix ${\bf P}_{h}=(p^{(h)}_{ij} )$ 
is defined as follows:
\[
p^{(h)}_{ij} = \left\{
\begin{array}{ll}
1 & \mbox{if $g_i h=g_j$,} \\
0 & \mbox{otherwise.}
\end{array}
\right.
\] 

Suppose that $p^{(h)}_{ij} =1 $, i.e., $g_j=g_ih$.
Then $U(e_{g_i}, f_{g_j } ) \neq 0$ if and only if $t(f,g_j)=o(e,g_i)$. 
Furthermore, $t(f,g_j)=o(e,g_i)$ if and only if 
$(o(e), g_{i} )=o(e, g_{i} )=t(f, g_{j} )=(t(f), g_{j} \alpha (f))$. 
Thus, $t(f)=o(e)$ and $ \alpha (f)=g^{-1}_{j} g_i =g^{-1}_{j} g_j h^{-1} =h^{-1} $. 
Similarly, $(f,g_j)=(e,g_i)^{-1} $ if and only if 
$f= e^{-1} $ and $ \alpha (f)= h^{-1} $. 
That is, under the assumption of (*), 
\[
U (e_g ,f_h ) =\left\{
\begin{array}{ll}
t_e ( x_{o(e) } - \delta {}_{e^{-1} f} ) & \mbox{if $t(f)=o(e)$ and $\alpha (f)= h^{-1} $, } \\
0 & \mbox{otherwise.}
\end{array}
\right.
\] 
Now, by (36),  
\[
{\bf U} ( G^{\alpha } )= \sum_{h \in \Gamma } {\bf P}_{h} \otimes {\bf U} {}_{h^{-1}} 
= \sum_{g \in \Gamma } {\bf P}_{g^{-1} } \otimes {\bf U} {}_{g} 
= \sum_{g \in \Gamma } {}^t {\bf P}_{g} \otimes {\bf U} {}_{g}   .
\] 
Here, note that ${\bf P}_{g^{-1} } = {}^t {\bf P}_{g} $ for each $g \in \Gamma $. 

Let $\rho$ be the right regular representation of $ \Gamma $.
Furthermore, let $ {\rho}_{1} =1, {\rho}_{2} , \ldots , {\rho}_{k} $
be all inequivalent irreducible representations of $ \Gamma $, and 
$f_i$ the degree of $ {\rho}_{i} $ for each $i$, where 
$f_1=1$.
Then we have $\rho (g)= {\bf P}_{g} $ for $g \in \Gamma $.
Furthermore, there exists a nonsingular matrix ${\bf P}$ such that
${\bf P}^{-1} \rho (g) {\bf P} = (1) \oplus f_2 \circ {\rho}_{2} (g) 
\oplus \cdots \oplus f_k \circ {\rho}_{k} (g)$ 
for each $g \in \Gamma $(see [30]). 
Thus, we have 
\[
{}^t {\bf P} {}^t \rho (g) {}^t {\bf P}^{-1}  = (1) \oplus f_2 \circ {}^t {\rho}_{2} (g) 
\oplus \cdots \oplus f_k \circ {}^t {\rho}_{k} (g) . 
\]

Putting 
${\bf F} =( {}^t {\bf P} \otimes {\bf I}_{2q} ) 
{\bf U} ( G^{\alpha } ) ( {}^t {\bf P}^{-1}  \otimes {\bf I}_{2q} )$,
we have 
\[
{\bf F}= \sum_{g \in \Gamma } 
\{ (1) \oplus f_2 \circ {}^t {\rho}_{2} (g) \oplus \cdots \oplus 
f_k \circ {}^t {\rho}_{k} (g) \} \otimes {\bf U} {}_g .
\]
Note that ${\bf U} (G)= \sum_{g \in \Gamma } {\bf U} {}_g $ and 
$1+ f^2_2 + \cdots + f^2_k =p$.
Therefore it follows that 
\[
\det ( \sigma {\bf I}_{2 mp} - {\bf U} ( G^{\alpha } ))= 
\det( \sigma {\bf I}_{2m} - {\bf U} (G)) 
\prod^{k}_{i=2} \det ( \sigma {\bf I}_{2m f_i } 
- \sum_{g} {}^t {\rho}_i (g) \otimes {\bf U} {}_g )^{f_i} . 
\]

Next, let $V(G)= \{ 1, \ldots , n \} $. 
Arrange vertices of $G^{ \alpha } $ in $p$ blocks:
$(1,1), \ldots , (n,1);$ 
$(1,g_2), \ldots , (n,g_2); 
\ldots ; (1,g_p), \ldots ,(n,g_p). $
We consider the matrix $\overline{{\bf A}} ( G {}^{ \alpha } ) $ 
defined in (***) under this order.

Suppose that $p^{(h)}_{ij} =1 $, i.e., $g_j=g_ih$.
Then $((u,g_i),(v,g_j)) \in D(G {}^{ \alpha } ) $
if and only if $(u,v) \in D(G)$ and 
$g_{j} = g_{i} \alpha (u,v)$.  
If $g_{j} = g_{i} \alpha (u,v)$, then 
$ \alpha (u,v)=g^{-1}_{i} g_j =g^{-1}_{i} g_i h=h$.
Thus we have
\[
\overline{{\bf A}} (G {}^{ \alpha } )= \sum_{h \in \Gamma } {\bf P}_{h} 
\otimes \overline{{\bf A}} {}_h .
\]

Putting 
${\bf E} =( {\bf P}^{-1} \otimes {\bf I}_p ) \overline{{\bf A}} 
(G {}^{ \alpha } ) ( {\bf P} \otimes {\bf I}_p )$ with nonsingular matrix ${\bf P} $, 
we have 
\[
{\bf E}= \sum_{h \in \Gamma } 
\{ (1) \oplus f_2 \circ {\rho}_{2} (h) \oplus \cdots \oplus 
f_k \circ {\rho}_{k} (h) \} \otimes \overline{{\bf A}} {}_h .
\]
Note that $\overline{{\bf A}} (G) = \sum_{h \in \Gamma } \overline{{\bf A}} {}_h $.  
Therefore it follows that 
\[
\begin{array}{rcl} 
\det ( {\bf I}_{np} - \sigma \overline{{\bf A}} (G^{\alpha } , \sigma {}^{2})
+ \overline{{\bf D}} ( G^{\alpha } , \sigma {}^{2}))
\  & = & \det ( {\bf I}_{n} - \sigma \overline{{\bf A}} + \overline{{\bf D}} ) \\
\  &   &                \\ 
\  & \times & \prod^{k}_{i=2} 
\det ( {\bf I}_{n f_i } - \sigma \sum_{h \in \Gamma } {\rho}_{i} (h) \otimes \overline{{\bf A}} {}_{h} 
+ {\bf I}_{f_i} \otimes \overline{{\bf D}} )^{f_i} .
\end{array}
\] 
Hence, it follows that 
\[
\det ( \sigma {\bf I}_{2 mp} - {\bf U} ( G^{\alpha } ))= \det( \sigma {\bf I}_{2m} - {\bf U} (G)) 
\prod^{k}_{i=2} \det ( \sigma {\bf I}_{2m f_i } - \sum_{h} {}^t {\rho}_i (h) \otimes {\bf U} {}_h )^{f_i} 
\]
\[
= \det ( {\bf I}_{n} - \sigma \overline{{\bf A}} + \overline{{\bf D}} )  
\prod^{k}_{i=2} \det ( {\bf I}_{n f_i } - \sigma \sum_{h \in \Gamma } {\rho}_{i} (h) \otimes \overline{{\bf A}} {}_{h} 
+ {\bf I}_{f_i} \otimes \overline{{\bf D}} )^{f_i} \prod^m_{j=1} (\sigma {}^2 - e^{2{\bf i}kL_{e_j} } )^p .  
\]

The third formula of Theorem is obtained similarly to the second one. 
\begin{flushright}$\Box$\end{flushright}
%%%%%%%%%%%%%%%%%%%%%%%%%%%%%%%%%%%%%%%%%%%%
%%%%%%%%%%%%%%%%%%%%%%%%%%%%%%%%%%%%%%%%%%%%

%%%%%%%%%%%%%%%%%%%%%%%%%%%%%%%%%%%%%%%%%%%%
%%%%%%%%%%%%%%%%%%%%%%%%%%%%%%%%%%%%%%%%%%%%
\section{$L$-functions of graphs}

We state a short review for the zeta function of a graph. 

A {\em path $P$ of length $n$} in $G$ is a sequence 
$P=( v_0, e_1, v_1 , e_2 ,v_2 , \ldots , v_{n-1} , e_n , v_n )$ of $n+1$ vertices 
and $n$ arcs such that $v_0 \in V(G)$, $v_i \in V(G)$, $ e_i \in D(G)$ 
and $ e_i =( v_{i-1} ,v_i )$ for $1 \leq i \leq n$. 
We write $P=(e_1, \ldots ,e_n )$.  
Set $|P|=n$, $o(P)= v_0$ and $t(P)= v_n $. 
Also, $P$ is called an {\em $(o(P),t(P))$-path}. 
We say that a path $P=(e_1, \ldots ,e_n )$ has a {\em backtracking} 
if $ e^{-1}_{i+1} =e_i $ for some $i$. 
A $(v, w)$-path is called a {\em $v$-cycle} 
(or {\em $v$-closed path}) if $v=w$. 
As standard terminologies of graph theory, a path and a cycle are 
a diwalk and a closed diwalk, respectively.

We introduce an equivalence relation on the set of cycles. 
Two cycles $C_1 =(e_1, \ldots ,e_m )$ and 
$C_2 =(f_1, \ldots ,f_m )$ are {\em equivalent} if there exists 
$k$ such that $f_j =e_{j+k} $ for all $j$. 
Let $[C]$ be the equivalence class that contains a cycle $C$. 
Let $B^r$ be the cycle obtained by going $r$ times around a cycle $B$. 
Such a cycle is called a {\em power} of $B$. 
A cycle $C$ is {\em reduced} if 
both $C$ and $C^2 $ have no backtracking. 
Furthermore, a cycle $C$ is {\em prime} if it is not a power of 
a strictly smaller cycle. 
Note that each equivalence class of prime, reduced cycles of a graph $G$ 
corresponds to a unique conjugacy class of 
the fundamental group $ \pi {}_1 (G,v)$ of $G$ at a vertex $v$ of $G$. 

The {\em Ihara zeta function} of a graph $G$ is defined to be 
a function of $u \in {\bf C}$ with $|u|$ sufficiently small, by 
\[
{\bf Z} (G, u)= {\bf Z}_G (u)= \prod_{[C]} (1- u^{ \mid C \mid } )^{-1} ,
\]
where $[C]$ runs over all equivalence classes of prime, reduced cycles 
of $G$ (see [18]).

\begin{theorem}[Bass]
If $G$ is a connected graph, then the reciprocal of the Ihara zeta function of $G$ 
is given by 
\[
{\bf Z} (G, u)^{-1} =(1- u^2 )^{r-1} \det ( {\bf I} -u {\bf A} (G)+ 
u^2 ({\bf D} -{\bf I} )), 
\]
where $r$ and ${\bf A} (G)$ are the Betti number and the adjacency matrix 
of $G$, respectively, and ${\bf D} = {\bf D}_G =( d_{ij} )$ is the diagonal matrix 
with $d_{ii} = \deg v_i $ where $V(G)= \{ v_1 , \ldots , v_n \} $. 
\end{theorem}

Stark and Terras [33] gave an elementary proof of Theorem 4 and 
discussed three different zeta functions of any graph. 
Other proofs of Bass' Theorem were given by 
Foata and Zeilberger [9] and Kotani and Sunada [23].

We introduce an $L$-function on the scattering matrix of a quantum graph.  
Let $G$ be a connected graph with $n$ vertices and $m$ unoriented edges, 
$ \Gamma $ be a finite group and $ \alpha : D(G) \longrightarrow \Gamma $ 
be an ordinary voltage assignment. 
Furthermore, let $ \rho $ be a unitary representation of $ \Gamma $ 
and $d$ its degree. 
We generalize the determinant of the second expression in Theorem 3.  
The {\em $L$-function} of $G$ associated with 
$ \rho $ and $ \alpha $ is defined by 
\[
\zeta {}_G (A,L, \lambda , \rho , \alpha , s)= \det ( {\bf I}_{2m d } 
-s \sum_{h \in \Gamma } {}^t {\rho} (h) \otimes {\bf U} {}_h )^{-1} . 
\]

If $\rho ={\bf 1}$ is the identity representation of $\Gamma $, then 
the reciprocal of the $L$-function of $G$ is a determinant on 
the bond scattering matrix of $G$. 

A determinant expression for the $L$-function of $G$ associated with 
$ \rho $ and $ \alpha $ is given as follows. 
For $1 \leq i,j \leq n$, the {\em $(i,j)$-block} ${\bf F}_{i,j} $ of a 
$dn \times dn$ matrix ${\bf F}$ is the submatrix of ${\bf F}$ 
consisting of $d(i-1)+1, \ldots , di$ rows and 
$d(j-1)+1, \ldots , dj$ columns.

\begin{theorem}
Let $G$ be a connected graph with $n$ vertices and $m$ unoriented edges, 
$ \Gamma $ be a finite group and 
$ \alpha : D(G) \longrightarrow \Gamma $ be an ordinary voltage assignment. 
If $ \rho $ is a representation of $ \Gamma $ and $d$ is 
the degree of $ \rho $, then the reciprocal of the $L$-function of $G$ 
associated with $ \rho $ and $ \alpha $ is
\[
\begin{array}{rcl}
\  &  &  \zeta {}_G (A,L, \lambda , \rho , \alpha ,s)^{-1} \\
\  &   &                \\ 
\  & = & 
\det ( {\bf I}_{nd} - s^{-1} \sum_{g \in \Gamma } \rho (g) \otimes \tilde{{\bf A}}_g ( s^{-2} )
+ {\bf I}_d \otimes \overline{{\bf D}} ( s^{-2} )) \prod^m_{j=1} (1- e^{2 {\bf i} kL_{e_j} } s^2 )^d \\
\  &   &                \\ 
\  & = & \det ( {\bf I}_{nd} - s^{-1} \sum_{g \in \Gamma } \rho (g) \otimes \overline{{\bf A}}_g ( s^{-2} )
+ {\bf I}_d \otimes \overline{{\bf D}} ( s^{-2} )) \prod^m_{j=1} (1- e^{2{\bf i} kL_{e_j} } s^2 )^d  ,  
\end{array}
\] 
where $D(G)= \{ e_1 , e^{-1}_1 , \ldots , e_m , e^{-1}_m \} $.
\end{theorem}

{\bf Proof}.  The argument is an analogue of the method of Watanabe and Fukumizu [40].  

Let $D(G)= \{ e_1, \ldots , e_m , e_{m+1},$ 
$\ldots , e_{2m} \} $ such that 
$ e_{m+i} = e^{-1}_i (1 \leq i \leq m)$. 
Furthermore, arrange arcs of $G$ as follows: 
\[
e_1 , e^{-1}_1 , \ldots , e_m , e^{-1}_m . 
\]  
Note that the $(e,f)$-block 
$(\sum_{g \in \Gamma } {\bf U}_g \otimes {}^t \rho (g) )_{ef} $ 
of $ \sum_{g \in \Gamma } {\bf U}_g \otimes {}^t \rho (g)$ is given by 
\[ 
(\sum_{g \in \Gamma } {\bf U}_g \bigotimes {}^t \rho (g) )_{ef} 
=\left\{
\begin{array}{ll}
{}^t \rho ( \alpha (f)) t_e ( x_{o(e) } - \delta {}_{e^{-1} f} ) & \mbox{if $t(f)=o(e)$, } \\
{\bf 0}_d  & \mbox{otherwise.}
\end{array}
\right.
\] 

For $ g \in \Gamma $, let the matrix ${\bf S}_g =( S^{(g)}_{ef} )$ 
be defined by
\[
S^{(g)}_{ef} =\left\{
\begin{array}{ll}
x_{o(e)} - \delta {}_{e^{-1} f} & \mbox{if $t(f)=o(e)$ and $ \alpha (f)=g$, } \\
0 & \mbox{otherwise.}
\end{array}
\right.
\] 
Then we have 
\[ 
(\sum_{g \in \Gamma } {\bf U}_g \otimes {}^t \rho (g) )  
=({\bf T} \otimes {\bf I}_d )
(\sum_{g \in \Gamma } {\bf S}_g \otimes {}^t \rho (g) ) . 
\]

For $g \in \Gamma $, two $2m \times 2m$ matrices 
${\bf B}_g =( {\bf B}^{(g)}_{ef} )_{e,f \in D(G)} $ and 
${\bf J}_g =( {\bf J}^{(g)}_{ef} )_{e,f \in D(G)} $ 
are defined as follows: 
\[
{\bf B}^{(g)}_{ef} =\left\{
\begin{array}{ll}
x_{t(e)} & \mbox{if $t(e)=o(f)$ and $ \alpha (e)=g$, } \\
0 & \mbox{otherwise}
\end{array}
\right.
, 
{\bf J}^{(g)}_{ef} =\left\{
\begin{array}{ll}
1 & \mbox{if $f= e^{-1} $ and $\alpha (e)=g$, } \\
0 & \mbox{otherwise.}
\end{array}
\right.
\]
Now  
\[
{}^t  {\bf S}_g = {\bf B}_g -{\bf J}_g \  for \  g \in \Gamma .  
\]
 
Let ${\bf K} =( {\bf K}_{ev} )$ ${}_{e \in D(G); v \in V(G)} $ be 
the $2md \times nd$ matrix defined 
as follows: 
\[
{\bf K}_{ev} :=\left\{
\begin{array}{ll}
{\bf I}_d & \mbox{if $o(e)=v$, } \\
{\bf 0}_d & \mbox{otherwise. } 
\end{array}
\right.
\]
Furthermore, we define a $2md \times nd$ matrix  
${\bf L} =( {\bf L}_{ev} ) {}_{e \in D(G); v \in V(G)} $ as follows: 
\[
{\bf L}_{ev} :=\left\{
\begin{array}{ll}
\rho ( \alpha (e)) & \mbox{if $t(e)=v$, } \\
{\bf 0}_d & \mbox{otherwise. } 
\end{array}
\right.
\]
Then we have  
\begin{equation}
{\bf L} ( {\bf X} \otimes {\bf I}_d ) {}^t {\bf K} = 
\sum_{h \in \Gamma } {\bf B} {}_h \otimes \rho (h)= {\bf B}_{\rho } ,    
\end{equation} 
where 
\[
{\bf B}_{\rho } = \sum_{g \in \Gamma } {\bf B}_g \otimes \rho (g) . 
\]

Now, let 
\[
{\bf X}_d = {\bf X} \otimes {\bf I}_d \  and \  
{\bf T}_d = {\bf T} \otimes {\bf I}_d . 
\]
Then we have  
\[
\begin{array}{rcl}
\  &   & \det ( {\bf I}_{2md} -s \sum_{g \in \Gamma } {}^t \rho (g) \otimes {\bf U}_g ) 
= \det ( {\bf I}_{2md} -s \sum_{g \in \Gamma } {\bf U}_g \otimes {}^t \rho (g) ) \\  
\  &   &                \\ 
\  & = & \det ( {\bf I}_{2md} -s {\bf T}_d ( \sum_{g \in \Gamma } {\bf S}_g \otimes {}^t \rho (g)) \\ 
\  &   &                \\ 
\  & = & \det ( {\bf I}_{2md} -s {\bf T}_d ( \sum_{g \in \Gamma } {}^t {\bf B}_g \otimes {}^t \rho (g) 
-\sum_{g \in \Gamma } {}^t {\bf J}_g \otimes {}^t \rho (g)) . 
\end{array}
\]
Set 
\[
{\bf J}_{\rho } = \sum_{g \in \Gamma } {\bf J}_g \otimes \rho (g) . 
\]
Thus, by (37),  
\[
\begin{array}{rcl}
\  &   & \det ( {\bf I}_{2md} -s \sum_{g \in \Gamma } {}^t \rho (g) \otimes {\bf U}_g )
= \det ( {\bf I}_{2md} -s {\bf T}_d ( {\bf K} {\bf X}_d {}^t {\bf L} - {}^t {\bf J}_{\rho } )) \\  
\  &   &                \\ 
\  & = & \det ( {\bf I}_{2md} +s {\bf T}_d {}^t {\bf J}_{\rho } -s {\bf T}_d {\bf K} {\bf X}_d {}^t {\bf L} ) \\ 
\  &   &                \\ 
\  & = & \det ( {\bf I}_{2md} -s {\bf T}_d {\bf K} {\bf X}_d {}^t {\bf L} 
( {\bf I}_{2md} +s {\bf T}_d {}^t {\bf J}_{\rho } )^{-1} ) 
\det ( {\bf I}_{2md} +s {\bf T}_d {}^t {\bf J}_{\rho }) . 
\end{array}
\]

By (33), we have 
\begin{equation} 
\begin{array}{rcl}
\  &   & \det ( {\bf I}_{2md} -s \sum_{g \in \Gamma } {}^t \rho (g) \otimes {\bf U}_g ) \\
\  &   &                \\ 
\  & = & \det ( {\bf I}_{nd} -s {\bf X}_d {}^t {\bf L} 
( {\bf I}_{2md} +s {\bf T}_d {}^t {\bf J}_{\rho } )^{-1} {\bf T}_d {\bf K} ) 
\det ( {\bf I}_{2md} +s {\bf T}_d {}^t {\bf J}_{\rho }) . 
\end{array}
\end{equation} 
Furthermore, 
\begin{equation} 
\begin{array}{rcl}
\  &   & \det ( {\bf I}_{2md} -s \sum_{g \in \Gamma } {}^t \rho (g) \otimes {\bf U}_g ) \\
\  &   &                \\ 
\  & = & \det ( {\bf I}_{nd} -s {}^t {\bf L} ( {\bf I}_{2md} +s {\bf T}_d {}^t {\bf J}_{\rho } )^{-1} 
{\bf T}_d {\bf K} {\bf X}_d )
\det ( {\bf I}_{2md} +s {\bf T}_d {}^t {\bf J}_{\rho }) . 
\end{array}
\end{equation} 

Next, we have 
\[
{\bf I}_{2md} +s {\bf T}_d {}^t {\bf J}_{ \rho } =
\left[ 
\begin{array}{ccccc}
{\bf I}_d & st_{e_1 } {}^t \rho ( \alpha (e^{-1}_1)) &   &   & {\bf 0} \\
st_{e^{-1}_1 } {}^t \rho ( \alpha (e_1 )) & {\bf I}_d &   &  \\ 
   &    & \ddots &   &   \\
   &    &   & {\bf I}_d & st_{e_m}  {}^t \rho ( \alpha (e^{-1}_m)) \\ 
{\bf 0} &    &   & st_{e^{-1}_m }  {}^t \rho ( \alpha (e_m )) & {\bf I}_d
\end{array} 
\right] 
,  
\]
and so, 
\[
\det ( {\bf I}_{2md} +s {\bf T}_d {}^t {\bf J}_{ \rho } ) 
= \prod^{m}_{j=1} (1- t_{e_j  } t_{e^{-1}_j } s^2 )^{d} . 
\]
Furthermore, we have 
\[
( {\bf I}_{2md} +s {\bf T}_d {}^t {\bf J}_{ \rho } )^{-1} 
= 
\left[ 
\begin{array}{ccc}
1/y_1 {\bf I}_d & - st_{e_1 } /y_1 {}^t \rho ( \alpha (e^{-1}_1)) & {\bf 0} \\ 
- st_{e^{-1}_1 } /y_1 {}^t \rho ( \alpha ( e_1 )) & 1/y_1 {\bf I}_d &  \\
{\bf 0}  &   &  \ddots 
\end{array} 
\right] 
,  
\]
where $y_j =1- t_{e_j} t_{e^{-1}_j }s^2 \  (1 \leq j \leq m)$.  

For an arc $(u,v) \in D(G)$, 
\[
( {\bf X}_d {}^t {\bf L} ( {\bf I}_{2md} +s {\bf T}_d {}^t {\bf J}_{\rho } )^{-1} 
{\bf T}_d {\bf K} )_{uv} 
= x_u t_{(v,u)} /(1- t_{(u,v)} t_{(v,u)} s^2 ) {}^t \rho ( \alpha (v,u)) . 
\]
Furthermore, if $u=v$, then 
\[
( {\bf X}_d {}^t {\bf L} ( {\bf I}_{2md} +s {\bf T}_d {}^t {\bf J}_{\rho } )^{-1} 
{\bf T}_d {\bf K} )_{uu} 
=- \sum_{t(e)=u} \frac{ x_u t_e t_{e^{-1}} s}{1- t_e t_{e^{-1} } s^2 } {\bf I}_d . 
\]
Then we have 
\[
{\bf X}_d {}^t {\bf L} ( {\bf I}_{2md} +s {\bf T}_d {}^t {\bf J}_{\rho } )^{-1} 
{\bf T}_d {\bf K} =1/s^2 \sum_{g \in \Gamma } {}^t \tilde{{\bf A}}_g (1/s^2 ) \otimes {}^t \rho (g) 
-1/s \overline{{\bf D}} (1/s^2 ) \otimes {\bf I}_d . 
\]
By (38), it follows that 
\[
\begin{array}{rcl}
\  &   & \det ( {\bf I}_{2md} -s \sum_{g \in \Gamma } {}^t \rho (g) \otimes {\bf U}_g ) \\  
\  &   &                \\ 
\  & = & \det ( {\bf I}_{nd} -s {\bf X}_d {}^t {\bf L} ( {\bf I}_{2md} 
+s {\bf T}_d {}^t {\bf J}_{\rho } )^{-1} {\bf T}_d {\bf K} ) \prod^{m}_{j=1} (1- t_{e_j  } t_{e^{-1}_j } s^2 )^{d} \\
\  &   &                \\ 
\  & = & \det ( {\bf I}_{nd} -1/s \sum_{g \in \Gamma } {}^t \tilde{{\bf A}}_g (1/s^2 ) \otimes {}^t \rho (g) 
+ \overline{{\bf D}} (1/s^2 ) \otimes {\bf I}_d  ) 
\prod^{m}_{j=1} (1- e^{2{\bf i} kL_{e_j} } s^2 )^{d} \\
\  &   &                \\ 
\  & = & \det ( {\bf I}_{nd} -1/s \sum_{g \in \Gamma } \tilde{{\bf A}}_g (1/s^2 ) \otimes \rho (g) 
+ \overline{{\bf D}} (1/s^2 ) \otimes {\bf I}_d  ) 
\prod^{m}_{j=1} (1- e^{2{\bf i} kL_{e_j} } s^2 )^{d} \\
\  &   &                \\ 
\  & = & \det ( {\bf I}_{nd} -1/s \sum_{g \in \Gamma } \rho (g) \otimes \tilde{{\bf A}}_g (1/s^2 ) 
+ {\bf I}_d \otimes \overline{{\bf D}} (1/s^2 )) \prod^{m}_{j=1} (1- e^{2{\bf i} kL_{e_j} } s^2 )^{d} . 
\end{array}
\]

Next, for an arc $(u,v) \in D(G)$, 
\[
( {}^t {\bf L} ( {\bf I}_{2md} +s {\bf T}_d {}^t {\bf J}_{\rho } )^{-1} 
{\bf T}_d {\bf K} {\bf X}_d )_{uv} 
= x_v t_{(v,u)} /(1- t_{(u,v)} t_{(v,u)} s^2 ) {}^t \rho ( \alpha (v,u)) . 
\]
Furthermore, if $u=v$, then 
\[
( {}^t {\bf L} ( {\bf I}_{2md} +s {\bf T}_d {}^t {\bf J}_{\rho } )^{-1} 
{\bf T}_d {\bf K} {\bf X}_d )_{uu} 
=- \sum_{o(e)=u} \frac{ x_u t_e t_{e^{-1}} s}{1- t_e t_{e^{-1} }s^2 } {\bf I}_d . 
\]
Then we have 
\[
{}^t {\bf L} ( {\bf I}_{2md} +s {\bf T}_d {}^t {\bf J}_{\rho } )^{-1} {\bf T}_d {\bf K} {\bf X}_d  
=1/s^2 \sum_{g \in \Gamma } {}^t \overline{{\bf A}}_g (1/s^2 ) \otimes {}^t \rho (g) 
-1/s \overline{{\bf D}} (1/s^2 ) \otimes {\bf I}_d . 
\]
By (39), it follows that 
\[
\begin{array}{rcl}
\  &   & \det ( {\bf I}_{2md} -s \sum_{g \in \Gamma } {}^t \rho (g) \otimes {\bf U}_g ) \\  
\  &   &                \\ 
\  & = & \det ( {\bf I}_{nd} -s {}^t {\bf L} ( {\bf I}_{2md} +s {\bf T}_d {}^t {\bf J}_{\rho } )^{-1} {\bf T}_d {\bf K} {\bf X}_d ) 
\prod^{m}_{j=1} (1- t_{e_j  } t_{e^{-1}_j } s^2 )^{d} \\
\  &   &                \\ 
\  & = & \det ( {\bf I}_{nd} -1/s \sum_{g \in \Gamma } {}^t \overline{{\bf A}}_g (1/s^2 ) \otimes {}^t \rho (g) 
+ \overline{{\bf D}} (1/s^2 ) \otimes {\bf I}_d  ) 
\prod^{m}_{j=1} (1- e^{2{\bf i} kL_{e_j} } s^2 )^{d} \\
\  &   &                \\ 
\  & = & \det ( {\bf I}_{nd} -1/s \sum_{g \in \Gamma } \overline{{\bf A}}_g (1/s^2 ) \otimes \rho (g) 
+ \overline{{\bf D}} (1/s^2 ) \otimes {\bf I}_d  ) 
\prod^{m}_{j=1} (1- e^{2{\bf i} kL_{e_j} } s^2 )^{d} \\
\  &   &                \\ 
\  & = & \det ( {\bf I}_{nd} -1/s \sum_{g \in \Gamma } \rho (g) \otimes \overline{{\bf A}}_g (1/s^2 ) 
+ {\bf I}_d \otimes \overline{{\bf D}} (1/s^2 )) \prod^{m}_{j=1} (1- e^{2{\bf i} kL_{e_j} } s^2 )^{d} . 
\end{array}
\]
$\Box$

Thus,

\newtheorem{corollary}{Corollary}
\begin{corollary}
Let $G$ be a connected graph with $n$ vertices and $m$ 
unoriented edges, $ \Gamma $ be a finite group and 
$ \alpha : D(G) \longrightarrow \Gamma  $ be an ordinary voltage assignment. 
If $ \rho $ is a irreducible representation of $ \Gamma $ and $d$ is 
the degree of $ \rho $, then 
\[
\begin{array}{rcl}
\  &  &  \det ( \sigma {\bf I}_{2md} - \sum_{h \in \Gamma } {}^t {\rho}_i (h) \otimes {\bf U} {}_h ) \\
\  &   &                \\ 
\  & = & 
\det ( {\bf I}_{nd} - \sigma \sum_{h \in \Gamma } {\rho} (h) \otimes \overline{{\bf A}} {}_{h} 
+ {\bf I}_{d} \otimes \overline{{\bf D}} ) \prod^m_{j=1} (\sigma {}^2 - e^{2{\bf i}kL_{e_j} } )^d \\
\  &   &                \\ 
\  & = & \det ( {\bf I}_{nd} -\sigma \sum_{h \in \Gamma } {\rho}_{i} (h) \otimes \tilde{{\bf A}} {}_{h} 
+ {\bf I}_{d} \otimes \overline{{\bf D}} ) \prod^m_{j=1} (\sigma {}^2 - e^{2{\bf i}kL_{e_j} } )^d ,     
\end{array}
\] 
where $D(G)= \{ e_1 , e^{-1}_1 , \ldots , e_m , e^{-1}_m \} $.
\end{corollary}

{\bf Proof}.  By Theorem 5, we have 
\[
\begin{array}{rcl}
\\  &   & \det ( \sigma {\bf I}_{2m f_i } - \sum_{h \in \Gamma } {}^t {\rho}_i (h) \otimes {\bf U} {}_h ) 
= \sigma {}^{2md} \zeta {}_G (A,L, \lambda , \rho , \alpha ,\sigma {}^{-1} )^{-1} \\
\  &   &                \\ 
\  & = & \det ( {\bf I}_{nd} - \sigma \sum_{g \in \Gamma } \rho (g) \otimes \tilde{{\bf A}}_g ( \sigma {}^2 )
+ {\bf I}_d \otimes \overline{{\bf D}} ( \sigma {}^2 )) 
\prod^m_{j=1} (\sigma {}^2 - e^{2 {\bf i} kL_{e_j} } )^d . 
\end{array}
\]
\begin{flushright}$\Box$\end{flushright}

By Theorem 5, it is also shown that, in Theorem 3, the determinant of the second expression is equal to 
that of the third expression. 

By Theorem 3 and Corollary 1, the following result holds.

\begin{corollary}
If $G$ is a connected graph with $m$ edges, $ \Gamma $ is a finite group and 
$ \alpha : D(G) \longrightarrow \Gamma $ is an ordinary voltage assignment,  
then we have 
\[
\det ( {\bf I}_{2 mp} - {\bf U} ( G^{\alpha }))= \sigma {}^{2mp} 
\prod_{ \rho } \zeta {}_G (A,L, \lambda , \rho , \alpha ,\sigma {}^{-1} )^{- \deg \rho } , 
\]
where $ \rho $ runs over all inequivalent irreducible representations 
of $ \Gamma $ and $p= \mid \Gamma \mid $. 
\end{corollary}

%%%%%%%%%%%%%%%%%%%%%%%%%%%%%%%%%%%%%%%%%%%%
%%%%%%%%%%%%%%%%%%%%%%%%%%%%%%%%%%%%%%%%%%%%

%%%%%%%%%%%%%%%%%%%%%%%%%%%%%%%%%%%%%%%%%%%%
%%%%%%%%%%%%%%%%%%%%%%%%%%%%%%%%%%%%%%%%%%%%
\section{The Euler product for the $L$-function 
$\zeta {}_G (A,L, \lambda , \rho , \alpha ,s)$ of a graph}

We present the Euler product for the $L$-function of a graph introduced in 
Section 6. 

Foata and Zeilberger [9] gave a new proof of Bass' Theorem by 
using the algebra of Lyndon words.
Let $X$ be a finite nonempty set, $<$ a total order in $X$, and 
$X^*$ the free monoid generated by $X$. 
Then the total order $<$ on $X$ derives the lexicographic order $<^*$ on $X^*$. 
A {\em Lyndon word} in $X$ is defined to a nonempty word in $X^*$ 
that is prime (not the power $l^r$ of any other word $l$ 
for any $r \geq 2$) and that is also minimal in the class of its 
cyclic rearrangements under $<^*$ (see [26]). 
Let $L$ denote the set of all Lyndon words in $X$. 

Foata and Zeilberger [9] gave a short proof of Amitsur's identity [3].

\begin{theorem}[Amitsur]
For square matrices ${\bf A}_1 , \ldots , {\bf A}_k $, 
\[
\det ( {\bf I} -( {\bf A}_1 + \cdots + {\bf A}_k ))= 
\prod_{l \in L} \det ( {\bf I} - {\bf A}_l ) , 
\]
where the product runs over all Lyndon words in $ \{ 1, \cdots , k \} $, 
and $ {\bf A}_l = {\bf A}_{i_1} \cdots {\bf A}_{i_r} $ for 
$l= i_1 \cdots i_r $. 
\end{theorem}

\begin{theorem}
Let $G$ be a connected graph with $n$ vertices and $m$ unoriented edges, 
$ \Gamma $ be a finite group and 
$ \alpha : D(G) \longrightarrow \Gamma $ be an ordinary voltage assignment. 
For each path $P=( e_1, \ldots , e_p )$ of $G$, set 
$ \alpha (P)= \alpha ( e_1 ) \cdots \alpha ( e_p )$. 
If $ \rho $ is a representation of $ \Gamma $ and $d$ is the degree of $ \rho $,  
then 
\[
\zeta {}_G (A,L, \lambda , \rho , \alpha ,s)= 
\prod_{[C]} \det ( {\bf I}_d - {}^t  \rho ( \alpha (C)) t_C a_C s^{|C|} )^{-1} , 
\]
where $[C]$ runs over all equivalence classes of prime cycles 
of $G$, and 
\[
t_C = t_{e_1} \cdots t_{e_p }, 
a_C = \sigma {}^{(o(e_1))}_{e_1 e_p } \sigma {}^{(o(e_{p} ))}_{e_p e_{p-1} } 
\cdots \sigma {}^{(o(e_2))}_{e_2 e_1 } , \  C=( e_1 , e_2 , \ldots , e_p ) 
\]
\end{theorem}

{\bf Proof}. At first, let 
$D(G)= \{ e_1 , \ldots , e_m , e_{m+1} , \ldots , e_{2m} \} $ and 
consider the lexicographic order on $D(G) \times D(G)$ derived from 
a total order of $D(G)$: $ e_1 < e_2 < \cdots < e_{2m} $. 
If $( e_i ,e_j )$ is the $c$-th pair under the above order, then 
we define the $2md \times 2md$ matrix 
${\bf T}_c =(( {\bf T}_c ) {}_{r,s} )_{1 \leq r,s \leq 2m }$ as follows: 
\[
( {\bf T}_c )_{r,s} =\left\{
\begin{array}{ll}
{}^t \rho ( \alpha ( e_j)) t_{e_i } \sigma {}^{(o(e_i)) }_{e_i e_j}  & \mbox{if $r=e_i, s=e_j$ and 
$o( e_i )=t(e_j )$, } \\
{\bf 0} & \mbox{otherwise, }
\end{array}
\right.
\]
where 
\[
\sigma {}^{(o(e))}_{ef}= x_{o(e) } - \delta {}_{e^{-1} f} . 
\]

If ${\bf F} ={\bf T}_1 + \cdots + {\bf T}_k $ and $k=4 m^2 $,  
then  
\[
{\bf F} = \sum_{h \in \Gamma } {\bf U} {}_h \otimes {}^t \rho (h) . 
\]

Let $L$ be the set of all Lyndon words in $D(G) \times D(G)$. 
We can also consider $L$ as the set of all Lyndon words in 
$ \{ 1, \ldots ,k \} $: 
$( e_{i_1 } ,e_{j_1 } ) \cdots ( e_{i_s } ,e_{j_s } )$ corresponds to 
$m_1 m_2 \cdots m_s $, where $( e_{i_r } ,e_{j_r } )(1 \leq r \leq s)$ 
is the $m_r $-th pair. 
Theorem 6 implies that 
\[
\det ( {\bf I}_{2qd} -s {\bf F} )= 
\prod_{t \in L} \det ( {\bf I}_{2md} -s^{|l|} {\bf T}_l ) , 
\]
where 
\[
{\bf T}_l ={\bf T}_{i_1} \cdots {\bf T}_{i_r} 
\] 
for $l= i_1 \cdots i_r $. 
Note that $ \det ( {\bf I}_{2md} -s^{|l|} {\bf T}_l )$ 
is the alternating sum of the diagonal minors of ${\bf T}_l $. 
Thus, we have 
\[
\det ( {\bf I} -s^{|t|}  {\bf T}_t ) =\left\{
\begin{array}{ll}
\det( {\bf I} - {}^t \rho ( \alpha (C)) t_C a_C s^{|C|} )  
& \mbox{if $t$ is a prime cycle $C$, } \\
1 & \mbox{otherwise, }
\end{array}
\right.
\] 
where 
\[
t_C = t_{e_1} \cdots t_{e_p }, 
a_C = \sigma {}^{(o(e_1))}_{e_1 e_p } \sigma {}^{(o(e_{p} ))}_{e_p e_{p-1} } 
\cdots \sigma {}^{(o(e_2))}_{e_2 e_1 } , \  C=( e_1 , e_2 , \ldots , e_p ) 
\]
Therefore, it follows that 
\[
\zeta {}_G (A,L, \lambda , \rho , \alpha ,s)^{-1} = 
\det ({\bf I}_{2md} -s \sum_{h \in \Gamma } {}^t \rho (h) \otimes {\bf U} {}_h ) 
\]
\[
= \det ({\bf I}_{2md} -s \sum_{h \in \Gamma } {\bf U} {}_h \otimes {}^t \rho (h)) 
= \prod_{[C]} ( {\bf I}_d - {}^t \rho ( \alpha (C)) t_C a_C s^{|C|} ) , 
\]
where $[C]$ runs over all equivalence classes of prime cycles 
of $G$. 
\begin{flushright}$\Box$\end{flushright}
%%%%%%%%%%%%%%%%%%%%%%%%%%%%%%%%%%%%%%%%%%%%
%%%%%%%%%%%%%%%%%%%%%%%%%%%%%%%%%%%%%%%%%%%%

%%%%%%%%%%%%%%%%%%%%%%%%%%%%%%%%%%%%%%%%%%%%
%%%%%%%%%%%%%%%%%%%%%%%%%%%%%%%%%%%%%%%%%%%%
\section{Example}

We give an example. See also Fig. 2, 
Let $G=K_3$ be the complete graph with three vertices $v_1 , v_2 , v_3 $ and 
six arcs $e_1 ,e_2 , e_3 , e^{-1}_1 ,e^{-1}_2 , e^{-1}_3 $, 
where $e_1 =( v_1 ,v_2 ), e_2 = ( v_2 ,v_3 ), e_3=( v_3 ,v_1 )$.  
Furthermore, let $\lambda {}_v = \lambda $ for $v \in V(G)$, 
$L_e =L , A_e =A$ for any $e \in D(G)$. 
Then we have 
\[
x_{v_j} = \frac{2{\bf i} k}{2{\bf i} k- \lambda } , 
t_{e_j } = \exp ({\bf i} L(k-A)) , 
t_{e^{-1}_j } = \exp ({\bf i} L(k+A)) \  (j=1,2,3) . 
\]
Set $a= \frac{2{\bf i} k}{2{\bf i} k- \lambda } $, $t= \exp ({\bf i} L(k-A))$ and 
$s= \exp ({\bf i} L(k+A))$. 
Considering ${\bf U} $ under the order 
$e_1 ,e_2 , e_3 , e^{-1}_1 ,e^{-1}_2 , e^{-1}_3 $, 
we have 
\[
{\bf U} = {\bf U}_{GS} =
\left[ 
\begin{array}{cccccc}
0 & 0 & ta & t(a-1) & 0& 0 \\
ta & 0 & 0 & 0 & t(a-1) & 0 \\ 
0 & ta & 0 & 0 & 0 & t(a-1) \\
s(a-1) & 0 & 0 & 0 & sa & 0 \\ 
0 & s(a-1) & 0 & 0 & 0 & sa \\
0 & 0 & s(a-1) & sa & 0 & 0 
\end{array} 
\right] 
\]
and  
\[
\tilde{{\bf A}} = \overline{{\bf A}} = \frac{a}{ \sigma {}^2 -st} 
\left[ 
\begin{array}{ccc}
0 & t & s \\
s & 0 & t \\ 
t & s & 0 
\end{array} 
\right] 
, 
{\bf D} = \frac{2ast}{\sigma {}^2 -st} {\bf I}_3 . 
\]
By Theorem 2, we have 
\[
\det (\sigma {\bf I}_6 - {\bf U} )=(\sigma {}^2 -st )^3 
\det ( {\bf I}_3 - \sigma \overline{{\bf A}} + \overline{{\bf D}} )
= \det  
\left[ 
\begin{array}{ccc}
b & -ta \sigma  & -sa \sigma  \\
-sa \sigma  & b & -ta \sigma  \\ 
-ta \sigma  & -sa \sigma & b 
\end{array} 
\right] 
\]
\[
=b( b^2 -3st a^2 \sigma {}^2 )- a^3 \sigma {}^3 ( s^3 +t^3 ).  
\]
where $b=\sigma {}^2 +(2a-1)st$. 

Next, let $\Gamma = Z_2 = \{ 1, -1 \} $ be 
the cyclic group of order 2, and let 
$ \alpha : D(K_3 ) \longrightarrow Z_2$ be the ordinary voltage 
assignment such that $ \alpha ( e_1 )= \alpha ( e^{-1}_1 )=-1$ 
and $ \alpha ( e_2 )= \alpha ( e^{-1}_2 )= \alpha ( e_3 )= \alpha (e^{-1}_3 )=1$. 
The characters of ${\bf Z}_2$ are given as follows: 
$\chi {}_i ((-1 )^j )=((-1)^i )^j $, $ 0 \leq i, j \leq 1$. 
Then we have 
\[
\overline{{\bf A}}_1 =\frac{a}{\sigma {}^2 -st} 
\left[ 
\begin{array}{ccc}
0 & 0 & s \\
0 & 0 & t \\ 
t & s & 0 
\end{array} 
\right] 
, 
\overline{{\bf A}}_{-1} =\frac{a}{\sigma {}^2 -st} 
\left[ 
\begin{array}{ccc}
0 & -s & 0 \\
-s & 0 & 0 \\ 
0 & 0 & 0 
\end{array} 
\right] 
.
\] 

Now, by Theorem 5, 
\[
\begin{array}{rcl}\sigma {}^6 {\zeta}_{K_3} (A,L, \lambda , \chi {}_1 , \alpha , \sigma {}^{-2} )^{-1} 
& = & (\sigma {}^2 -st )^3 \det ( {\bf I}_3 - \sigma \sum^1_{i=0} 
\chi ((-1 )^i ) \overline{{\bf A}}_{ (-1 )^i } + \overline{{\bf D}} ) \\ 
\  &   &                \\ 
\  & = &  
\det  
\left[ 
\begin{array}{ccc}
b & ta \sigma & -sa \sigma \\
sa \sigma & b & -ta \sigma \\ 
-ta \sigma & -sa \sigma & b  
\end{array} 
\right] 
\\ 
\  &   &                \\ 
\  & = & b( b^2 -3st a^2 \sigma {}^2 )+ a^3 \sigma {}^3 ( s^3 +t^3 ).  
\end{array}
\]

By Corollary 2, it follows that 
\[
\begin{array}{rcl}
\  &   & \det (\sigma {\bf I}_{12} - {\bf U} ( K^{\alpha }_3 ))= 
\det ({\sigma {}^2 \bf I}_6 - {\bf U} ) \zeta {}_{K_3} (A,L,\lambda , \chi , \alpha , \sigma {}^{-2} )^{-1} 
\sigma {}^6 \\ 
\  &   &                \\ 
\  & = & \{ b( b^2 -3st a^2 \sigma {}^2 )- a^3 \sigma {}^3 ( s^3 +t^3 ) \} 
\{ b( b^2 -3st a^2 \sigma {}^2 )+ a^3 \sigma {}^3 ( s^3 +t^3 ) \} \\
\  &   &                \\ 
\  & = &
b^2 ( b^2 -3st a^2 \sigma {}^2 )^2 - a^6 \sigma {}^6 ( s^3 +t^3 )^2 . 
\end{array}
\]

%%%%%%%%%%%%%%%%%%%%%%%%%%%
\begin{figure}
\begin{center}
\includegraphics[width=10cm]{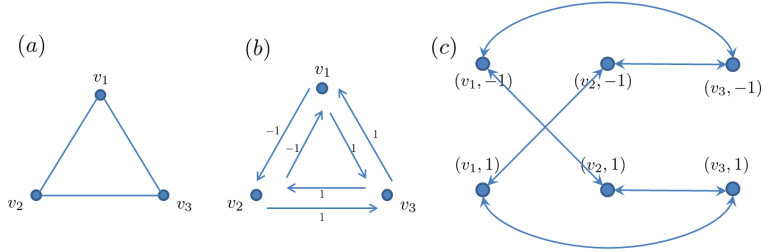}
\caption{ {\scriptsize Regular covering of $G=K_3$: 
Figure (a) is the original graph $G=K_3$. We take $\Gamma=\mathbb{Z}_2=\{-1,1\}$. 
As an ordinary assignment $\alpha$, we assign elements of $\Gamma$ to arcs as is depicted by Fig.~(b). 
For example, $\alpha(1,2)=-1$, $\alpha(2,3)=1$. The definition of the assignment 
imposes $\alpha(2,1)=-1$, $\alpha(3,2)=1$ since $\alpha(v,u)=\alpha(u,v)^{-1}$ for any $(u,v)\in D(G)$. 
Figure (c) is the derived graph $G^\alpha$. 
For example, putting $e=(v_1,v_2)\in D(G)$, then, for $\pm 1\in \Gamma$, $e_1=((v_1,1),(v_2,-1))\in D(G^\alpha)$ and $e_{-1}=((v_1,-1),(v_2,1))\in D(G^\alpha)$
since $((u,g),(v,h))\in D(G^\alpha)$ if and only if $(u,v)\in D(G)$ and $h=g \alpha(u,v)$, in this case, 
$-1=\alpha(v_1,v_2)\times 1$ and $1=\alpha(v_1,v_2)\times (-1)$, respectively.
}
}
\end{center}
\end{figure}
%%%%%%%%%%%%%%%%%%%%%%%%%%%

\vspace{5mm} 

{\bf Acknowledgments.}

The first author was supported in part by the Grant-in-Aid for Scientific
Research (C) 20540133 and (B) 24340031 from Japan Society for the Promotion of
Science. The second and third authors also acknowledge financial supports of the Grant-in-
Aid for Scientific Research (C) from Japan Society for the Promotion of Science (Grant No.
24540116 and No. 23540176, respectively).

%Acknowledgments.

\end{document}